\documentclass[useAMS,usenatbib]{mn2e}
\bibpunct{[}{]}{,}{n}{}{;}

\usepackage{epsfig,graphicx,times,subfigure}
\newcommand{\kms}{{km~s$^{-1}$}}
\newcommand{\teff}{{$T_\mathrm{eff}$}}
\newcommand{\logg}{{log~{\em g}}}

\newcommand{\vt}{{$\xi_\mathrm{t}$}}
\newcommand{\vsini}{$v_{\rm e}\sin{i}$}
\newcommand{\sini}{sin\,{\em i}}
\newcommand{\msun}{M$_{\odot}$}

\newcommand{\ew}{W$_{\lambda}$}

\newcommand{\aaps}{A\&A}
\newcommand{\aj}{AJ}
\newcommand{\mnras}{MNRAS}
\newcommand{\apj}{ApJ}
\newcommand{\apjl}{ApJL}

\def\deg{\hbox{$^\circ$}}

\title[Rotational velocities of Galactic B-type supergiants]
{Atmospheric parameters and rotational velocities for a sample of Galactic B-type
supergiants} 
\author[M. Fraser et al.]
{M. Fraser$^{1}$\thanks{E-mail:mfraser02@qub.ac.uk},
P.L. Dufton$^{1}$, I. Hunter$^{1}$, R.S.I. Ryans$^{1}$\\
$^{1}$Department of Physics and Astronomy, Queen's University of Belfast,
Belfast BT7 1NN, Northern Ireland.
\\
}
\begin{document}

\date{Submitted to Monthly Notices of the Royal Astronomical Society}

\pagerange{\pageref{firstpage}--\pageref{lastpage}} \pubyear{}

\maketitle

\label{firstpage}

\begin{abstract}

High resolution optical spectra of 57 Galactic B-type supergiant stars have been analyzed to
determine their rotational and macroturbulent velocities. In addition, their
atmospheric parameters (effective temperature, surface gravity and
microturbulent velocity) and surface nitrogen abundances have been estimated 
using a non-LTE grid of model atmospheres. Comparisons of the
projected rotational velocities have been made with the predictions of stellar
evolutionary models and in general good agreement was found. However for a small
number of targets, their observed rotational velocities were significantly larger
than predicted, although their nitrogen abundances were consistent with the rest
of the sample.  We conclude that binarity may have played a role in generating
their large rotational velocities. No correlation was found between nitrogen
abundances and the current projected rotational velocities. However a
correlation was found with the inferred projected rotational velocities of the
main sequence precursors of our supergiant sample. This correlation is again in
agreement with the predictions of single star evolutionary models that
incorporate rotational mixing. The origin of the macroturbulent and 
microturbulent velocity fields is discussed and our results support  previous
theoretical studies that link the former to sub-photospheric convection and the
latter to non-radial gravity mode oscillations. In addition, we have attempted to
identify differential rotation in our most rapidly rotating targets.

\end{abstract}


\section{Introduction}

The spectra of early-type stars are affected by a variety of non-thermal 
velocity fields. Those present in the photosphere are often
designated as microturbulence (see, for example, Gies \& Lambert
\citealp{Gie92}; Daflon, Cunha \& Butler \citealp{Daf04}; Sim\'{o}n-D\'{i}az et al.
\citealp{Sim06}) and macroturbulence (see, for example, Howarth et al.
\citealp{How97}; Ryans et al. \citealp{Rya02}). Microturbulence is associated
with distance scales smaller than the mean free path of a photon, while macroturbulence
is associated with longer distance scales. Recently, the former has been
linked with the presence of sub-surface convection fields (Cantiello et al.
\citealp{Can09}) driven by the opacity of iron group elements, whilst the latter
may be a manifestation of the large number of  non-radial gravity-mode stellar
oscillations (Aerts et al. \citealp{Aer09}) present in the stellar photosphere.

Stellar rotation also significantly affects early-type stellar spectra and
particularly for main sequence objects can dominate the broadening of the metal
absorption lines (see, for example, Gray \citealp{Gra76}, \citealp{Gra08}). Rotation may play a
key role in the evolution of early-type stars (see, for example, Maeder \&
Maynet \citealp{Mae01}) and in long-duration gamma-ray bursts (Woosley \& Heger
\citealp{Woo06}). In turn this has stimulated several recent studies (Abt,
Levato \& Grosso \citealp{Abt02}; Huang \& Gies  \citealp{Hua06}; Hunter et al.
\citealp{Hun08a}; Martayan et al. \citealp{Mar06}; Strom, Wolff \& Dror \citealp{Str05}; Wolff et al. 
\citealp{Wol82, Wol07}) of stellar rotation in both field and cluster stars and
in different metallicity environments. However recently both slowly
rotating stars with relatively high nitrogen abundance and rapidly rotating
stars with relatively low nitrogen abundances (Hunter et al. 
\citealp{Hun08b, Hun09})
have been identified in the  VLT-FLAMES survey of massive stars 
(Evans et al. \citealp{Eva05}). Hence other mechanisms as well as rotation 
may be important for mixing nucleosynthetically processed material to the 
surface, as discussed by Brott et al. (\citealp{Bro08}). 

In this paper, we present high resolution and high signal-to-noise 
spectra of a sample of 57 Galactic B-type supergiant stars, and use these data to
further investigate the relationship between rotation and mixing, in addition to
testing the predictions of theoretical models. We also characterize the
microturbulent and macroturbulent velocity fields present in our sample and
compare these with the theoretical studies of Cantiello et al. (\citealp{Can09})
and Aerts et al. (\citealp{Aer09}) respectively. We have also attempted to make the
first identification of differential rotation in evolved massive stars.

\section{Observations and data reduction} \label{Obs_red}

The observing list was initially developed to sample as widely as possible
B-type supergiants of different spectral subtype. The availibility of targets
(in terms of spatial position and brightness) led to the early-B spectral types being
better sampled than the later spectral types as can be seen from 
Table\,\ref{Obs_data}. However even for the  latter, most
spectral subclasses contained at least one target. A number of  targets were
subsequently excluded from the analysis. In particular, four B9  supergiants
(HD\,94367, HD\,96919, HD\,111904, HD\,158799) were found to  have effective
temperatures that were smaller than the lower limit of our  model atmosphere
grid. Additionally eight targets  (HD\,74804, HD\,99103, HD\,75821, HD\,76728,
HD\,101330, HD\,152046, HD\,68161, HD\,53244) were  found to have  gravities
that were significantly higher than supegiants of similar spectral  types.
Indeed their logarithmic gravity estimates (3.0-3.5 dex) were similar to  those
found for B-type giants (see, for example, Vrancken et al. \citealp{Vra00}) and
they were therefore excluded.

\begin{figure}
\includegraphics[angle=270,width=3.5in]{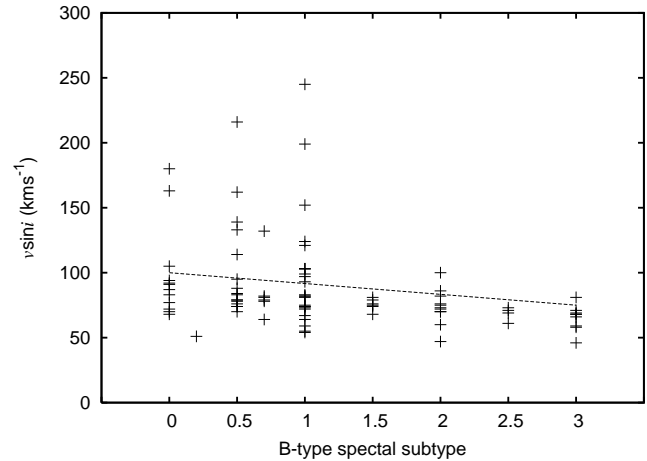}
\caption{Estimated projected rotational velocities as a function of 
spectral type for the early B-type supergiants analysed by
Howarth et al. (\citealp{How97}). The dashed line represents a subjective
estimate of the upper limit for the part of the sample that are rotating 
relatively slowly.}
\label{Howarth}
\end{figure}

Previous investigations of B-type supergiants (Ryans et al.
\citealp{Rya02};  Dufton et al. \citealp{Duf06}; Sim{\'o}n-D{\'{\i}}az \&
Herrero \citealp{Sim07})  have shown that many have relatively small
projected rotational velocities. In order to try to increase our sample of rapidly
rotating supergiants (in order  to, for example, investigate whether they
have a different evolutionary history), we have used the projected
rotational velocities estimated by Howarth et al. (\citealp{How97}) from
IUE spectra, as a guide in selecting additional targets. These were determined by a cross-correlation technique, which
implicity assumed that the spectral line broadening was dominated by
rotation. Howarth et al. discussed the validity of this assumption and
inferred from the lack of supergiants with very small projected rotational
velocities that another important broadening mechanism was probably
present. Indeed for the majority of their B-type supergiants this mechanism
probably dominates as illustrated by Fig. \ref{Howarth}, which plots
their estimated projected rotational velocity against spectral type. Most of 
the sample lie on a locus with an upper bound
stretching from  approximately 100 \kms\ at a B0 spectral type to 75 \kms\ at
B3. For these  targets the projected rotational velocities are probably
small and the broadening is dominated by macroturbulence. This is supported
by fig. 2 of Dufton et al. (\citealp{Duf06}), where a similar dependence of
the macroturbulent broadening with spectral type was found. However several
supergiants have estimates that appear to be too large for their
spectral type and in these cases the total line  broadening is probably
being significantly augmented by rotation. Four of these targets had already
been included in our observing lists but six additional targets had
positions and apparant magnitudes that made them accessible and these were
observed and have been identified with an asterisk in the
Table\,\ref{Obs_data}.

High resolution (R$\simeq$48,000) spectra were obtained with the Fibre
fed  Extended Range Optical Spectrograph (FEROS; Kaufer et al.
\citealp{Kau99}) on the 2.2-m MPI/ESO telescope at La Silla during a three
night observing run in April 2005. Some preliminary observations had been
obtained in July and  December 2004 to test the feasibility of our
analysis procedures  (see Sect. \ref{data_anal}) but due to their
relatively low signal-to-noise-ratios (SNRs) these have not been used in
the current analysis. All spectra covered the  wavelength range from
3600 \AA\ to 9200 \AA\ and were reduced using the  reduction pipeline that
runs under the MIDAS environment (Kaufer et al. \citealp{Kau99}).

The signal-to-noise ratio (SNR) for each spectrum was estimated by taking a
normalised region of the continuum and finding the standard deviation, the
SNR estimate being its reciprocal. This procedure was normally carried out
in the region of 4500-4600 \AA\ as this was compatible with the spectral
lines used to estimate the projected rotational velocities (see Table\,
\ref{Lines_used}). However for spectral types later than B5, this region 
contained a significant number of strong lines and hence the region  from
4700-4800 \AA\ was adopted. As the SNR is expected to vary with 
wavelength, the estimates for several stars were measured in both   regions
with those from the 4700-4800 \AA\ region being found to be  approximately 5 per cent
lower. Hence the estimates for all stars of spectral  type B6 or later have
been scaled accordingly. Given that weak undetected  absorption will lead to
underestimates of the SNRs, these estimates should  only be taken as an
approximate comparative guide to the quality of the  spectra. All the
targets are listed in the Table \ref{Obs_data}, 
together with exposure times and estimates of the SNRs. Spectral  types
are taken in the first instance from the Bright Star Catalog (Hoffleir, \citealp{Hof91}),
then from the IUE Atlas of B-Type Stellar Spectra (Walborn, Parker \& Nichols,
\citealp{Wal95}), the University of Michigan Catalogue of Spectral Types
(Houl \& Cowley, \citealp{Hou75}) and Reed (\citealp{Ree03}).

For all stars in the sample, the equivalent widths (\ew) of selected lines 
were estimated. The N II line at 3995 \AA\ was measured to estimate nitrogen
abundances, whilst the silicon spectrum (Si II 4128 \AA, 4131 \AA;  Si III
4552 \AA, 4567 \AA, 4574 \AA; and Si IV 4116 \AA) was used to constrain effective
temperatures as described in Section \ref{ap}.  The continuum adjacent to
each line was normalised using a second-order  polynomial fit.  For the
slowly rotating stars (see Sect.\ \ref{Vsini}) where macroturbulent 
broadening will generally dominate, a Gaussian was then fitted
to the absorption line using a least squares technique. For those  stars with
a relatively large projected rotational velocity  ($\approx100$ \kms\ or
greater), the lineshape is dominated by rotation. In these cases, a profile
shape appropriate to rotationally dominated broadening (with the value of 
\vsini\ taken from Table \ref{Vsini_atm} - see Sect\ \ref{Vsini}) was used, 
and was generally found to give a better fit. Although the choice of
profile  is a possible source of error, we note that the equivalent width
estimates of the stronger lines normally agreed to within 5 to 10 m\AA\ with
those found by direct numerical integration over the profile. The estimates
are presented  in Table \ref{eq_widths}.

\begin{table}
\begin{center}
\caption[]{Observational details.}
\begin{tabular}{llrll}
\hline\hline
Identifier & Sp.\ Type & Exp.\ (s) & SNR & Ref.
\\\hline

HD\,77581$^*$   &	B0 Ia	        		&	716   &	430 	&  4$^{\dagger}$
\\
HD\,122879	&	B0 Ia          	 &	464	&	510 	& 1
\\
HD\,149038	&	B0 Ia           	&	119   &	470  	& 1
\\
HD\,167264	&	B0 Iab		&	178	&	500	& 1
\\
HD\,164402	&	B0 Ib			&	258	&	550	& 1$^{\dagger}$
\\
HD\,168021	&	B0 Ib			&	684	&	400 	& 1$^{\dagger}$ 
\\	
HD\,115842	&	B0.5 Ia		&	321	&	480	& 1
\\	
HD\,150898	&	B0.5 Ia		&	214	&	500 	& 1
\\	
HD\,152234	&	B0.5 Ia		&	190	&	450	& 1$^{\dagger}$ 
\\	
HD\,152667$^*$	&	B0.5 Ia	&	362	&	350	& 1
\\	
HD\,94493$^*$	&	B0.5 Iab		&	1010	&	520	& 3
\\	
HD\,64760       &	B0.5 Ib		&	62	&	480	& 1
\\	
HD\,103779	&	B0.5 II		&	962	&	500	& 3
\\	
HD\,155985	&	B0.7 Ib		&	486	&	480	& 2		
\\
HD\,109867	&	B1 Ia			&	397	&	460	& 1		
\\	
HD\,152235	&	B1 Ia			&	424	&	420	& 1 		
\\																		
HD\,142758	&	B1 Ia			&	830	&	410	& 3
\\	
HD\,148688	&	B1 Iae		&	170	&	380	& 1
\\	
HD\,154090	&	B1 Ia			&	111	&	470	& 1
\\	
HD\,152236	&	B1 Iape		&	98	&	440	& 1
\\	
HD\,119646      &	B1 Ib-II		&	543	&	410	& 3
\\	
HD\,125545	&	B1 Iab-Ib		&	1145	&	400	& 3
\\	
HD\,150168	&	B1 Iab-Ib		&	229	&	390	& 1
\\	
HD\,99857$^*$	&	B1 Ib			&	1220	&	400 	& 3
\\	
HD\,157246	&	B1 Ib			&	27	&	560	& 1
\\
HD\,93840$^*$	&	BN1 Ib		&	1640	&	410	& 2 		
\\																	
HD\,106343	&	B1.5 Ia		&	386	&	400	& 1
\\		
HD\,148379	&	B1.5 Iape		&	173	&	400	& 1		
\\	
HD\,96248	&	BC1.5 Iab		&	528	&	420	& 2		
\\		
HD\,108002	&	B2 Ia-Iab		&	750	&	400	& 3   		
\\																		
HD\,52089 	&	B2 II			&	5	&	430	& 1
\\		
HD\,99953	&	B2 Iab-Ib		&	495	&	440	& 3
\\	
HD\,93827$^*$	&	B2 Ib	-II		&	6770	&	270	& 3
\\	
HD\,111990	&	B2 Ib			&	647	&	400	& 3 
\\		
HD\,117024	&	B2 Ib			&	861	&	400	& 3
\\		
HD\,165024	&	B2 Ib			&	37	&	510	& 1
\\		
HD\,141318	&	B2 II			&	246	&	450  & 1
\\		
HD\,92964	&	B2.5 Iae		&	178	&	440	& 1
\\		
HD\,116084	&	B2.5 Ib		&	270	&	500	& 1
\\																
HD\,75149	&	B3 Ia			&	192	&	440	& 1 		
\\		
HD\,53138	&	B3 Iab		&	20	&	500	& 1
\\		
HD\,51309   	&	B3 II			&	70	&	520	& 1
\\																	
HD\,157038	&	B4 Ia			&	460	&	450	& 1 
\\			
HD\,159110	&	B4 Ib			&	1045	&	550	& 4
\\		
HD\,79186	&	B5 Ia			&	126	&	440	& 1
\\		
HD\,111973	&	B5 Ia			&	288	&	480 	& 1 		
\\		
HD\,58350	&	B5 Ia			&	12	&	440	& 1
\\		
HD\,86440	&	B5 Ib			&	33	&	520	& 1
\\		
HD\,164353	&	B5 Ib			&	49	&	520	& 1
\\		
HD\,83183	&	B5 II			&	54	&	400	& 1 
\\
HD\,74371	&	B6 Iae		&	155	&	440	& 1
\\		
HD\,80558	&	B6 Iae		&	280	&	430	& 1
\\		
HD\,105071  	&	B6 Iab-Ib		&	428	&	510	& 1	
\\		
HD\,125288	&	B6 Ib			&	68	&	500	& 1
\\																	
HD\,91619	&	B7 Iae		&	359	&	460	& 1		
\\											
HD\,111558	&	B7 Ib		&	1007	&	450	& 3 		
\\		
HD\,166937	&	B8 Iap		&	44	&	460	& 1$^{\dagger}$
	
\\		
\hline	
\multicolumn{5}{l}{\(^{*}\) Subsequently added to observing list as 
Howarth et al. (\citealp{How97})}
\\ 
\multicolumn{5}{l}{found anomalously large line widths for spectral type}
\\
\multicolumn{5}{l}{$^{\dagger}$ Listed as binary in SIMBAD online database.}
\\
\multicolumn{5}{l}{(1) Hoffleit \& Warren (\citealp{Hof91}), (2) Walborn, Parker \& Nichols (\citealp{Wal95}), }
\\ 
\multicolumn{5}{l}{(3) Houk \& Cowley (\citealp{Hou75}), (4) Reed (\citealp{Ree03})}
\\ 
\label{Obs_data}
\end{tabular}
\end{center}
\end{table}

\section{Data analysis} 
\label{data_anal}

\subsection{Projected rotational velocities} 
\label{Vsini}

The use of Fourier transforms to estimate projected 
rotational velocities (\vsini) from stellar spectra was first proposed by Carroll 
(\citealp{Car33}), and more recently has been discussed by Gray
(\citealp{Gra76}, \citealp{Gra08}), Reiners \& Schmitt (\citealp{Rei02}) and Royer
(\citealp{Roy05}). The applicability of this methodology specifically to
OB-type stars has been considered by Sim{\'{o}}n-D{\'{\i}}az et al.
(\citealp{Sim06}). In essence the Fourier Transform technique
is predicated on the convolution theorem, whereby when an
observed spectrum is transformed into the Fourier domain, the convolution
of the intrinsic spectrum with the rotational, macroturbulent and
instrumental profiles becomes a multiplication of the corresponding
Fourier transforms.  Furthermore, as only the rotational profile is
expected to have zeroes in its Fourier Transform at low frequencies, these
will appear in the total transform, thereby allowing the projected
rotational velocity to be estimated.

The position of the first zero for a solidly rotating body is inversely 
proportional to the projected rotational velocity (see, for example,  Gray
\citealp{Gra76}, \citealp{Gra08}); this relationship will form the basis for our estimates
of the projected rotational velocities. A more  sophisticated analysis, such
as that discussed by Reiners (\citealp{Rei03}) shows that differential
rotation, limb darkening, the angle of inclination and gravity darkening can
all alter the line profile, and hence its Fourier transform. However, these
effects should not significantly affect the position of the first zero,
particularly for stars with an equatorial velocity, \( v <200 \) \kms\, which
is appropriate to most of our sample. Furthermore, the main  diagnostic of
these secondary effects -- namely the ratio of the first and second zeroes
in the Fourier transform, (\(q_{1}/{q_2}\)) -- was only measurable for
stars which had significant rotational broadening. Such objects will have
large equatorial velocities and are discussed further in Sect.\,\ref{r_diffrot}.

As discussed in Sect.\ \ref{Obs_red}, there is additional broadening present
in the spectrum of early-type supergiants (Howarth et al. \citealp{How97};
Ryans et. al. \citealp{Rya02}; Sim{\'{o}}n-D{\'{\i}}az et. al.
\citealp{Sim06}), which is normally designated as macroturbulence. This will
also affect the line profile in the Fourier domain, acting to decrease the
amplitude of the sidelobes. While the precise nature of this broadening is
not fully understood, if it can be approximated by, for example, a Gaussian 
it should not 
affect the postion of the first zero in the Fourier Transform. Recently Aerts et
al. (\citealp{Aer09}) simulated the effects of non-radial gravity mode
oscillations on the observed line profile. In particular, they find that on occasions this
can lead to an underestimation of the projected rotational velocity and we
discuss this further in Sect. \ref{results}.

Thirteen metal spectral lines were considered for each star, as listed in Table
\ref{Lines_used}. Four of the lines used are in close doublets or triplets with
separations of 0.25 \AA\ or less (corresponding to a velocity of approximately
15 \kms), which is less than the typical broadening from macroturbulence and 
rotation. Indeed, in all these cases, the features appeared as single lines. Hydrogen and
diffuse helium lines, although strong and well observed, were not normally
considered, as although the Fourier technique should be capable  of separating
rotational broadening from that due to linear Stark effect, the approximation of
line broadening as a convolution of unrelated effects may become unreliable
(Heinzel, \citealp{Hei78}). An exception was made for two targets (HD\,64760
and HD\,152667) with large projected rotational velocities, where the
metal lines were poorly observed and/or blended; for these targets, the
diffuse neutral helium lines at 3819\AA\ and 4026\AA\ were also considered.

Spectra were analysed using {\sc procspec}, an unpublished package of
{\sc idl} routines for manipulating spectra. 
The lines listed in Table \ref{Lines_used} were
normalised using a polynomial that had been fitted to the adjacent 
continuum. Whilst the amount of continuum considered should be minimised to
reduce the noise in the Fourier domain, tests showed that the projected
rotational velocity estimates were not  significantly affected by this
factor.

\begin{figure}
\includegraphics[angle=0,width=3.2in]{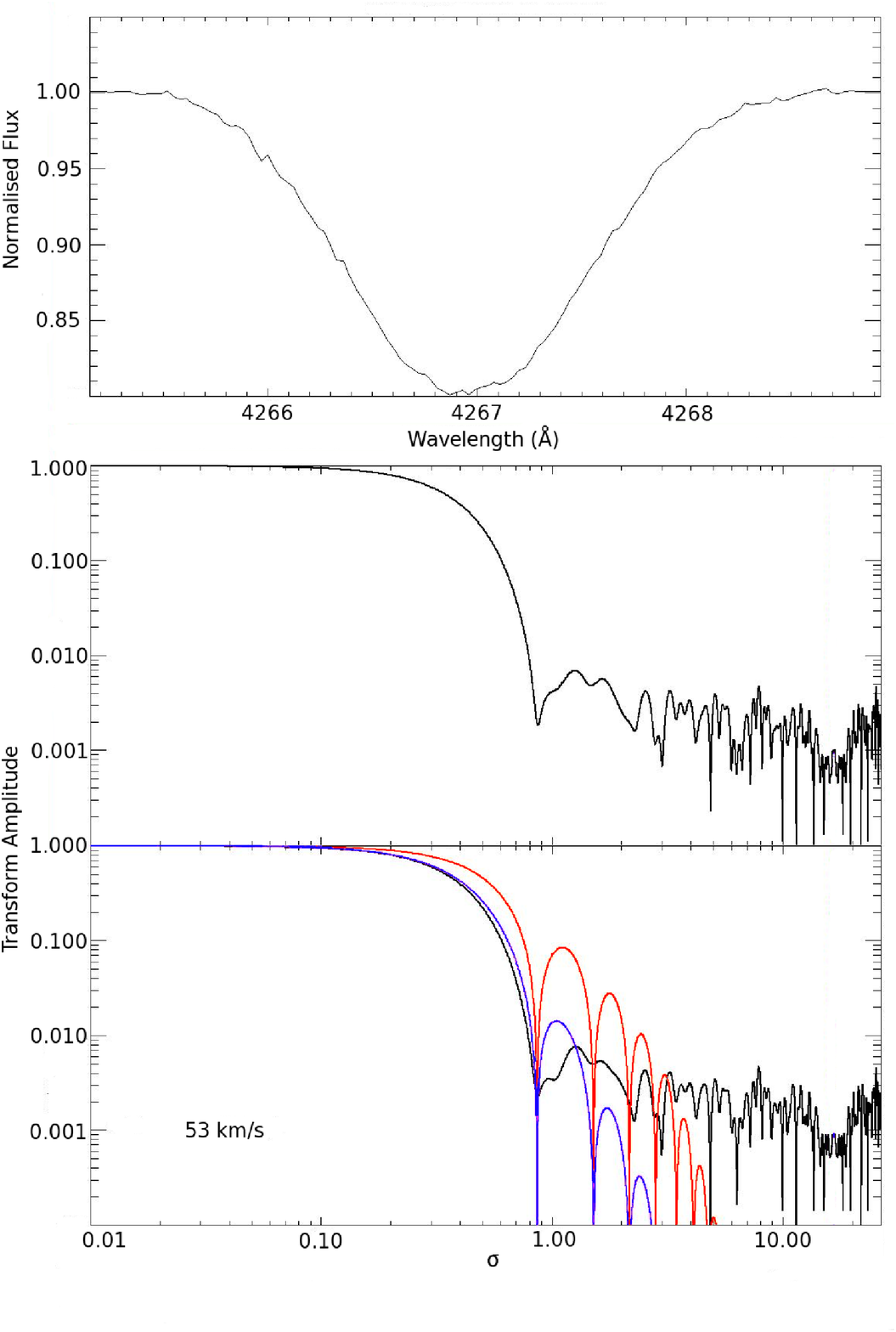}
\caption{An example of the Fourier transform methodology for the CII 
4267\AA\ line in HD117024. The topmost panel shows the observed
lineprofile, normalised to the continuum. The centre panel 
shows the Fourier transform of this line. The bottom panel shows the 
same FT, but with two theoretical Fourier transforms overplotted, one with no 
macroturbulence (red line), and one with a Gaussian macroturbulent velocity 
of 50\kms (blue line). The value of \vsini\ which gave the best fit to this line 
was 53 \kms.}
\label{example_fig1}
\end{figure}

Each line was then assigned a ranking between 1 and 5 to reflect its
reliability  (relative to the other lines for that star). Factors considered
were the quality of the spectrum, degree of symmetry in the line, the
presence of cosmic rays, confidence in the normalisation procedure, and the
intrinsic strength of the line. Lines that were in the top two quality
ranking bins for each star were Fourier transformed using the procedure
discussed by Sim{\'{o}}n-D{\'{\i}}az \& Herrero (\citealp{Sim07}).
Those that gave a transform with a clearly identifiable first zero were 
then used to estimate the projected rotational velocity. An illustrative 
example of this procedure can be seen in Fig. \ref{example_fig1}.
For each star, the number, mean and standard deviations of these estimates of 
the projected rotational velocity are summarized in Table \ref{Vsini_atm}.

\subsection{Macroturbulence} \label{macroturbulence}
Macroturbulence was originally introduced as an additional velocity field to
improve the agreement between observed metal line profiles  of supergiants and
those calculated using model atmospheres (see, for example, Howarth et al.
\citealp{How97};  Sim{\'{o}}n-D{\'{\i}}az et al. \citealp{Sim06}; Ryans et al. \citealp{Rya02}).  Macroturbulence is postulated
to be a velocity field characterised by a length scale longer than the mean free
path of a photon (as opposed to microturbulence). Recent work by Aerts et al. 
(\citealp{Aer09})  has suggested a physical explanation for
macroturbulence, in term of  the pulsation of stars due to non-radial
gravity-mode oscillations.  These authors considered pulsational models for a
B1-type supergiant, with comparable atmospheric parameters to those  found in
this paper and found that the additional line broadening simulated a
macroturbulence consistent with those found observationally. Additionally the
pulsations could lead to errors in the values estimated for  projected
rotational velocities and we return to this in Sect. \ref{r_macro}.

We have estimated macroturbulences for all the stars in our sample by assuming
that the magnitude of the velocity field follows a Gaussian distribution.
Clearly given the work of Aerts et al., this assumption may not be valid.
However the estimates will still be useful in characterising how the excess
broadening varies with spectral type and in many cases where macroturbulence
dominates the line broadening, the observed profiles appeared to be well 
approximated by a Gaussian function. 

Our procedure was to consider the lines tabulated in Table \ref{Lines_used} 
(excluding the neutral helium line) and to optimise the agreement between
theoretical and observed profiles. Intrinsic spectra were extracted from our
{\sc tlusty} grid using the grid point closest to our estimated atmospheric
parameters and an elemental abundance that gave an equivalent width similar to
that observed - typically differences were less than ten per cent. The theoretical
profile was then  scaled so that the two equivalent widths were the same. The
theoretical profile was convolved with a rotational broadening function using
the values of the projected rotational velocity listed in Table \ref{Vsini_atm}
and then with Gaussian profiles to represent the macroturbulence. For each line
an estimate of the macroturbulence was found by minimising the sum of the square
of the residuals with the mean, standard deviation and number of the estimates
being listed in Table \ref{Vsini_atm}. Note that we have
defined the macroturbulence as the  $\frac{1}{e}$\  (half-) width of the Gaussian
profile. These values are discussed further in Sect.\ \ref{r_macro}.

\begin{table}
\caption[]{Lines used to estimate the projected rotational velocities and
macroturbulence for our targets.}
\begin{center}
\begin{tabular}{lc}\hline\hline
Species   & Wavelength (\AA)
\\\hline
N II & 3995.00
\\
Si II & 4128.07
\\
Si II & 4130.89
\\	
C II & 4267.00	/ 4267.26
\\
Mg II & 4481.13 / 4481.33
\\
Si III & 4552.62
\\
Si III & 4567.82
\\	
Si III & 4574.76
\\
O II & 4590.97
\\
O II & 4595.96 / 4596.18
\\
N II & 4630.54
\\
O II & 4661.63
\\
He I & 4713.15 / 4713.38
\\

\hline
\label{Lines_used}
\end{tabular}
\end{center}
\end{table}

\begin{table*}
\begin{center}
\caption[]{Estimates of the projected rotational velocity, atmospheric
parameters and nitrogen abundances. The ordering of
targets follows that in Table\ref{Obs_data}}
\begin{tabular}{lcccccccclcl}
\hline\hline
Identifier & \vsini & $\sigma_r$ & n$_r$ & v$_t$ & $\sigma_t$ & n$_t$ & \teff & \logg & \vt\ & [$\frac{N}{H}$]& M \\
& \kms & \kms & & \kms & \kms & & K & cm s$^{-2}$ & \kms & & \msun
\\\hline
HD\,77581$^{++}$  & 56	 & 10 & 4  &  -  &  - &  - & 26500              & 2.90 & 17 & 8.40 & 40$^{**}$ \\
HD\,122879 & 55  &  3 & 6  &  80 &  7 &  6 & 27200              & 2.90 & 14 & 8.37 & 40$^{**}$ \\
HD\,149038 & 61  &  3 & 5  &  69 &  8 &  5 & 28500              & 3.10 & 12 & 7.91 & 40$^{**}$ \\      
HD\,167264 & 83	 &  7 & 5  &  66 &  7 &  6 & 27500              & 3.05 & 16 & 7.96 & 30$^{**}$ \\      
HD\,164402$^{++}$ & 46	 &  6 & 7  &  56 &  7 &  7 & 28100              & 3.25 & 13 & 7.55 & 30$^{**}$ \\ 
HD\,168021$^{++}$  & 48	 &  9 & 7  &  58 &  7 &  7 & 26900              & 3.10 & 15 & 7.81 & 25$^{**}$ \\     
HD\,115842 & 39	 &  3 & 6  &  66 &  5 &  6 & 24800              & 2.75 & 14 & 8.44 & 40 \\     
HD\,150898 & 93  &  4 & 4  &  56 &  9 &  6 & 26400              & 3.10 & 20 & 8.00 & 25$^{**}$ \\	
HD\,152234$^{++}$ & 59  &  3 & 4  &  57 &  3 &  8 & 25700              & 2.90 & 14 & 7.69 & 38$^{**}$ \\	
HD\,152667 & 139 &  5 & 8  &   - &  - &  - & 26000$^{\dagger}$  & 3.05 & 18 & 7.58 & 25$^{**}$ \\
HD\,94493  & 97  &  3 & 7  &  60 &  5 &  8 & 23500              & 2.95 & 14 & 8.02 & 24$^{**}$ \\	
HD\,64760  & 255 &  7 & 5  &   - &  - &  - & 26000$^{\dagger}$  & 3.25 & 22 & 7.89 & 23 \\	
HD\,103779 & 40  &  1 & 6  &  60 &  4 &  6 & 25000              & 3.00 & 12 & 8.07 & 25$^{**}$ \\	
HD\,155985 & 35	 &  4 & 9  &  46 &  3 &  8 & 23200              & 2.95 & 12 & 8.19 & 22 \\      
HD\,109867 & 43  &  5 & 9  &  56 &  4 &  9 & 22300              & 2.75 & 14 & 8.32 & 23 \\
HD\,152235 & 46  &  3 & 7  &  58 &  7 &  8 & 22000              & 2.65 & 14 & 7.59 & 30 \\																	
HD\,142758 & 46	 &  8 & 6  &  38 &  7 &  7 & 17300              & 2.25 & 21 & 8.46 & 35 \\	 
HD\,148688 & 48	 &  6 & 7  &  44 &  6 &  9 & 20700              & 2.45 & 16 & 7.94 & 37 \\	 
HD\,154090 & 47	 &  5 & 4  &  54 &  4 &  8 & 22200              & 2.70 & 15 & 8.30 & 28 \\	 
HD\,152236 & 34	 &  3 & 5  &  46 &  6 &  8 & 21500$^*$          & -    & -  & -    & -  \\	 
HD\,119646 & 35  &  5 & 6  &  26 &  4 &  8 & 17600              & 2.50 & 21 & 7.93 & 20 \\	
HD\,125545 & 65  &  5 & 7  &  53 &  4 &  9 & 20800              & 2.75 & 17 & 7.70 & 22 \\	
HD\,150168 & 129 &  7 & 5  &  55 & 12 &  6 & 24800              & 3.15 & 13 & 8.26 & 20$^{**}$ \\	
HD\,99857  & 189 & 10 & 6  &  59 & 11 &  7 & 21500$^{\dagger}$  & 2.95 & 16 & 7.20 & 18 \\
HD\,157246 & 269 & 11 & 4  &   - &  - &  - & 21500$^{\dagger}$  & 2.90 & 15 & 8.11 & 19 \\																	
HD\,93840  & 58  &  3 & 6  &  68 &  4 &  6 & 20900              & 2.75 & 15 & 8.37 & 23 \\	
HD\,106343 & 43  &  4 & 8  &  52 &  4 &  7 & 20100              & 2.50 & 17 & 8.20 & 30 \\	
HD\,148379 & 44  &  5 & 8  &  33 &  4 &  8 & 17000              & 2.00 & 19 & 7.93 & 43 \\																		  
HD\,96248  & 37	 &  5 & 7  &  53 &  4 &  8 & 19500              & 2.40 & 15 & 7.48 & 29 \\	 
HD\,108002 & 37  &  4 & 8  &  46 &  4 &  8 & 20200              & 2.60 & 16 & 8.19 & 25 \\	
HD\,52089  & 22  &  3 & 6  &  17 &  3 &  5 & 20100              & 3.05 & 17 & 7.74 & 14 \\		  
HD\,99953  & 49  &  3 & 5  &  37 &  4 &  6 & 16800              & 2.15 & 22 & 8.43 & 32 \\	  
HD\,93827  & 227 & 14 & 4  &   - &  - &  - & 18500$^{\dagger}$  & 2.70 & 20 & 7.86 & 17 \\		  
HD\,111990 & 34  &  6 & 7  &  30 &  5 &  9 & 16500              & 2.40 & 18 & 8.12 & 20 \\		  
HD\,117024 & 55  &  2 & 9  &  25 &  4 &  9 & 16500              & 2.55 & 24 & 7.62 & 12 \\		  
HD\,165024 & 98  &  2 & 7  &  35 &  7 &  9 & 18500              & 2.70 & 18 & 8.23 & 17 \\		  
HD\,141318 & 30  &  3 & 8  &  32 &  3 &  8 & 18300              & 2.90 & 15 & 8.13 & 13 \\		  
HD\,92964  & 36  &  4 & 8  &  28 &  4 &  8 & 15600              & 2.00 & 22 & 8.60 & 30 \\		  
HD\,116084 & 40  &  3 & 6  &  35 &  5 &  8 & 16200              & 2.25 & 23 & 8.09 & 23 \\	
HD\,75149  & 30  &  5 & 6  &  34 &  4 &  5 & 15900              & 2.20 & 20 & 8.43 & 24 \\		
HD\,53138  & 35  &  4 & 7  &  23 &  7 &  6 & 15400              & 2.15 & 18 & 8.22 & 24 \\		
HD\,51309  & 22  &  4 & 5  &  20 &  3 &  5 & 15600              & 2.40 & 18 & 8.23 & 17 \\																		
HD\,157038 & 38	 &  5 & 4  &  41 &  4 &  7 & 15700              & 2.00 & 14 & 8.97 & 32 \\		
HD\,159110 & 13	 &  4 & 9  &   9 &  3 &  8 & 19700              & 3.20 & 12 & 7.36 & 12 \\
HD\,79186  & 39 &   5 & 10 &  36 &  4 &  8 & 15100              & 2.00 & 14 & 8.39 & 35 \\	
HD\,111973 & 36  &  4 & 7  &  28 &  4 &  7 & 16000              & 2.30 & 19 & 8.04 & 21 \\		
HD\,58350  & 32 &   7 &	9  &  25 &  4 &  7 & 14500              & 2.10 & 18 & 8.16 & 23 \\		
HD\,86440  & 17 &   5 & 4  &  15 &  2 &  4 & 14600              & 2.55 & 20 & 7.59 & 13 \\	
HD\,164353 & 18 &   3 & 6  &  14 &  6 &  6 & 15100              & 2.50 & 21 & 7.78 & 14 \\		
HD\,83183  & 19 &   4 & 6  &   - &  - &  - & 14900              & 2.50 & 20 & 7.42 & 14 \\		
HD\,74371  & 31  & 4  & 5  &  30 &  3 &  5 & 13400              & 1.90 & 20     & 7.85 & 24\\	
HD\,80558  & 41  & 4  & 6  &  17 &  5 &  5 & 13000              & 1.70 & 24     & 8.66 & 30\\
HD\,105071 & 29  & 8  & 4  &  17 &  3 &  4 & 12200              & 1.85 & 20$^+$ & 7.97 & 20\\
HD\,125288 & 23  & 7  & 3  &  21 &  5 &  4 & 13900              & 2.55 & 20$^+$ & 7.67 & 10\\
HD\,91619  & 31  & 4  & 6  &  25 &  5 &  6 & 13100              & 1.70 & 18     & 8.36 & 31\\						
HD\,111558 & 26  & 10 & 4  &  21 &  9 &  4 & 12100              & 2.00 & 20     & 8.08 & 17\\
HD\,166937$^{++}$ & 29  & 5  & 4  &  32 &  2 &  4 & 12200              & 1.75 & 20     & 7.87 & 23\\
\hline	
\multicolumn{12}{l}{\(^{\dagger}\) large \vsini\ precludes use of silicon 
ionization balance and \vsini\ estimates supplemented 
with those from He lines}
\\
\multicolumn{12}{l}{\(^{*}\) gravity is too low to be analysed with
{\sc tlusty} grid}
\\
\multicolumn{12}{l}{\(^{**}\) In ambiguous region of HR diagram that 
can lead to ambiguous mass estimates}
\\
\multicolumn{12}{l}{\(^{+}\) Si\,{\sc iii} lines too weak to reliably 
estimate the microturbulence}
\\
\multicolumn{12}{l}{\(^{++}\) Classified as binary in SIMBAD}
\label{Vsini_atm}
\end{tabular}
\end{center}

\end{table*}

\begin{center}
\begin{table*}
\caption[]{Equivalent widths (in m\AA) for selected metal lines. 
The ordering of targets follows Table \ref{Obs_data}. The letter
B indicates a blend of lines.}
\begin{tabular}{lcccccccccccccc}
\\

\hline\hline
Identifier   & N II 3995 & Si IV 4116 & Si II 4129 & Si II 4131 & Si III 4553 & Si III 4568 & Si III 4575
\\\hline
HD\,77581    & 155      & 475	    & 	         &	      & 435	   & 365        & 195 \\
HD\,122879   & 105      & 455	    &            &	      & 305	   & 230        & 115 \\
HD\,149038   & 60       & 460	    &            & 	      & 230	   & 175        & 85  \\
HD\,167264   & 75       & 460	    & 	         &	      & 270	   & 200        & 95  \\
HD\,164402   & 50       & 345	    &            & 	      & 270	   & 210        & 110 \\
HD\,168021   & 85       & 365	    &            & 	      & 300	   & 245        & 125 \\
HD\,115842   & 165      & 395	    &            & 	      & 405	   & 345        & 195 \\
HD\,150898   & 155      & 370	    &            & 	      & 325	   & 250        & 125 \\
HD\,152234   & 80       & 380	    &            & 	      & 350	   & 295        & 160 \\
HD\,152667   & 85       & B	    & 	         &	      & 495  	   & 400        & 230 \\
HD\,94493    & 205      & 190	    & 	         &	      & 375	   & 310        & 180 \\
HD\,64760    & 170      & B	    & 	         &	      & 410	   & 330        & 155 \\
HD\,103779   & 165      & 255	    &            & 	      & 345	   & 285        & 165 \\
HD\,155985   & 235      & 165	    & 	         &	      & 375	   & 315        & 195 \\
HD\,109867   & 280      & 180	    & 	         &	      & 425	   & 355        & 215 \\
HD\,152235   & 125      & 230	    & 	         & 	      & 495	   & 425        & 270 \\
HD\,142758   & 385      & 85	    & 	         &	      & 555	   & 470        & 305 \\
HD\,148688   & 210      & 200	    & 	         &	      & 550	   & 480        & 305 \\
HD\,154090   & 285      & 190	    & 	         &	      & 470	   & 400        & 245 \\
HD\,152236   & 470      & 95	    & 	         &	      & 500	   & 435        & 275 \\
HD\,119646   & 235      & 40	    & 	         &	      & 380	   & 315        & 185 \\
HD\,125545   & 200      & 135	    & 	         &	      & 505	   & 425        & 270 \\
HD\,150168   & 230      & 230	    & 	         &	      & 385	   & 330        & 195 \\
HD\,99857    & 95       & B         & 	         &	      & 405	   & 330        & 200 \\
HD\,157246   & 290      & B         & 	         &	      & 460	   & 375        & 255 \\
HD\,93840    & 350      & 105	    & 	         &	      & 420	   & 345        & 215 \\
HD\,106343   & 330      & 115	    & 	         &	      & 465	   & 385        & 235 \\
HD\,148379   & 240      & 40	    & 	         &	      & 430	   & 365        & 230 \\
HD\,96248    & 140      & 95	    & 	         &	      & 465	   & 395        & 240 \\
HD\,108002   & 315      & 115	    & 	         &	      & 450	   & 380        & 235 \\
HD\,52089    & 200      & 55	    & 	         &	      & 350	   & 295        & 190 \\
HD\,99953    & 375      & 60	    & 	         &	      & 460	   & 380        & 235 \\
HD\,93827    & 225      & B	    & 	         &	      & 390	   & 325        & 210 \\
HD\,111990   & 215      & 20	    & 	         &	      & 275	   & 215        & 130 \\
HD\,117024   & 130      & 20	    & 	         &	      & 330	   & 270        & 160 \\
HD\,165024   & 300      & 55	    & 	         &	      & 405	   & 340        & 215 \\
HD\,141318   & 220      & 30	    & 	         &	      & 310	   & 250        & 160 \\
HD\,92964    & 345      & 25	    & 	         &	      & 365	   & 300        & 180 \\
HD\,116084   & 240      & 30	    &            &	      & 370	   & 305        & 180 \\
HD\,75149    & 285      & 	    & 140	 & 170        & 255	   & 200        & 115 \\
HD\,53138    & 200      & 	    & 155	 & 180	      & 230  	   & 175        & 105 \\
HD\,51309    & 120      & 	    & 150        & 160        & 165	   & 125        & 70  \\
HD\,157038   & 340      & 	    & 170	 & 190        & 300	   & 250        & 155 \\
HD\,159110   &	95      & 	    & 65         & 65	      & 195	   & 150        & 90  \\
HD\,79186    & 195      & 	    & 155        & 170        & 210	   & 160        & 95  \\
HD\,111973   & 190      & 	    & 150	 & 170        & 230 	   & 180        & 105 \\
HD\,58350    & 150      & 	    & 190        & 195        & 155	   & 115        & 65  \\
HD\,86440    & 60       & 	    & 165        & 175        & 85	   & 55         & 25  \\
HD\,164353   & 100      & 	    & 175        & 185        & 130	   & 90         & 50  \\
HD\,83183    & 55       & 	    & 175        & 185        & 75	   & 45         & 20  \\
HD\,74371    & 80       & 	    & 205        & 215        & 115	   & 80         & 45  \\
HD\,80558    & 105      & 	    & 200        & 205	      & 95  	   & 65	        & 35  \\
HD\,105071   & 55       & 	    & 250        & 255        & 65	   & 35         & 15  \\
HD\,125288   & 50       & 	    & 170        & 175        & 60	   & 30         & 15  \\
HD\,91619    & 135      & 	    & 200        & 210        & 110	   & 80         & 45  \\
HD\,111558   & 55       & 	    & 305        & 315	      & 50	   & 30         & 15  \\
HD\,166937   & 50       & 	    & 250        & 260        & 60	   & 40         & 20  \\
\hline

\label{eq_widths}
\end{tabular}
\end{table*}
\end{center}

\section{Atmospheric parameters} \label{ap}

Atmospheric parameters and nitrogen abundances have been estimated from
non-LTE model atmosphere grids generated using the codes {\sc tlusty} and  
{\sc synspec} (Hubeny \citealp{Hub88}; Hubeny \& Lanz \citealp{Hub95}; 
Hubeny et al. \citealp{Hub98}; Lanz and Hubeny \citealp{Lan07}). Details
of the methods adopted can be found in Hunter et al. (\citealp{Hun07})
while a more detailed discussion of the grids  can be found in Ryans et
al. (\citealp{Rya03}) and Dufton et al. (\citealp{Duf05})\footnote{See 
also http://star.pst.qub.ac.uk}. Hence only a brief summary will be given
here with further details available as discussed above.

Four model atmosphere grids have been calculated for metallicities
correponding to a Galactic metallicity of [Fe/H]=7.5 dex,  and metallicities
of 7.2, 6.8 and 6.4 dex to represent the LMC, SMC and lower metallicity
material respectively.  For each of these grids non-LTE models have been
calculated for effective temperatures ranging from 12\,000K to 35\,000K, in
steps of no more than 2\,500K, surface gravities ranging from 4.5\,dex down
to the Eddington limit, in steps of no more than 0.25\,dex and for
microturbulences of 0, 5, 10, 20, and 30\,km\,s$^{-1}$. Assuming that the
light elements (C, N, O, Mg and Si) have a negligible effect on
line-blanketing and the structure of the stellar atmosphere, models were
then generated with the light element abundance varied by +0.8, +0.4, -0.4
and -0.8\,dex about their normal abundance at each point on the {\sc tlusty}
grid. Theoretical spectra and equivalent widths were then calculated based
on these models. Photospheric abundance estimates for approximately 200 
absorption lines at any set of atmospheric parameters covered by the grid 
can then be calculated by interpolation between the models via simple {\sc idl}
routines. The reliability of the interpolation technique has been verified by
Ryans et al. (\citealp{Rya03}).

As the estimation of the atmospheric parameters is inter-related, it 
was necessary to use an iterative approach, as follows:  
\\
\\
- initial estimates of the effective temperature (\teff) and gravity
(\logg) were obtained from the spectral type and the calibration of Crowther,
Lennon \& Walborn (\citealp{Cro06}).
\\
- the microturbulence (\vt) was estimated from the Si\,{\sc iii} triplet 
at 4552--4574\AA.
\\
- the effective temperature was re-estimated from the silicon 
ionization equilibrium.
\\
- surface gravity was re-estimated from the Balmer line profiles.
\\
- the last three steps were iterated until the estimates converged.
\\
\\
Further details of this approach are given below.

\subsection{Effective temperature} \label{s_teff}

For the hotter targets with spectral types earlier than B3, the  Si\,{\sc
iii} to Si\,{\sc iv} ionization balance was used, whilst for the later
spectral types the Si\,{\sc ii} to Si\,{\sc iii} ionization  balance was
considered. The very high quality of the observational data implies that the
random errors should be small and normally less than 1\,000 K.
Systematic errors due to, for example, errors in  the adopted atomic data or
in the physical assumptions were more difficult to quantify. However, Dufton
et al. (\citealp{Duf05}) compared atmospheric parameters found for SMC
supergiants using the approach adopted here with those found for two stars 
by Trundle et al. (\citealp{Tru04}). The latter utilized the unified code {\sc fastwind}
(Santolaya-Rey, Puls \& Herrero, A. \citealp{San97}; Herrero, Puls \&
Najarro \citealp{Her02}; Repolust, Puls \& Herrero \citealp{Rep04}). The two
approaches yielded similar estimates of the effective temperature (agreeing
to 500K for one star and 2000K for the other), hence an error estimate of 
$\pm$1\,000 K would appear to be appropriate. The values of the other 
atmospheric parameters from the two codes were also in good agreement;
\logg\ to within 0.1 dex, microturbulence to within 3 \kms and nitrogen abundance
to within 0.2 dex.

Five targets (HD\,152667, HD\,64760, HD\,99857, HD\,157246, HD\,93827)
had large projected rotational velocities that precluded the observation
of two ionization stages of silicon. In these cases, the effective temperature was taken to
be that implied by the spectral type and then the other atmospheric parameters
were estimated as discussed below.

\subsection{Surface Gravity}          \label{s_logg}

The logarithmic surface gravity (\logg) of each star was estimated by 
fitting the observed hydrogen Balmer lines with theoretical profiles. In
order to minimise the effects of the stellar wind, relative high order
lines in the series (H$\gamma$ and H$\delta$) were normally considered. 
Automated procedures  have been developed to fit model spectra in our 
{\sc tlusty} model  atmosphere grid to the observed spectra, with contour 
maps displaying the region of best fit. With an effective temperature 
estimate being available (from the methods described above), it was a 
straightforward matter to estimate  the gravity. 

Hunter et al. (\citealp{Hun07}) have discussed the errors associated with
this procedure and estimate a typical random error of $\pm$0.1 dex. Given 
that our observational data is generally of a higher quality than their 
Magellanic Cloud spectroscopy, this would appear to be a conservative error
estimate to adopt here. Systematic errors due to, for example, uncertainties
in the line broadening or assumption in the model atmosphere calculations
could be significant, although the gravity estimates from the {\sc tlusty} grid 
and {\sc fastwind} calculations discussed in Dufton et al. (\citealp{Duf05}) 
agreed to within 0.1 dex. Hence an error estimate of $\pm$0.2 dex would therefore
appear to be a conservative estimate.

\subsection{Microturbulence}    \label{s_xi}

The microturbulence is normally derived by removing any systematic 
dependence of the abundance estimates on line strength found from lines of a specific ion. 
For B-type stars, the O\,{\sc ii} ion  is often considered (for example,
Sim\'{o}n-D\'{i}az et al. \citealp{Sim06};  Hunter et al.
\citealp{Hun05};  Gies \& Lambert \citealp{Gie92}; and Daflon et al.
\citealp{Daf04}) as its rich spectrum should improve its  reliablity.
However the analysis is complicated by the lines arising from  different
multiplets, making any estimate susceptible to errors in the adopted
atomic data or in  the magnitude of non-LTE effects. In order to remove
these uncertainties, a single O\,{\sc ii} multiplet can be used. However
the number of lines is then drastically reduced, whilst the O\,{\sc ii} spectrum is 
weak or unobservable for spectral types later than B3.

In order to maintain consistency throughout the analysis we have instead
estimated the microturbulence from  the Si\,{\sc iii} triplet of lines at
4552--4574 \AA, which is observed in all our spectra. As 
these lines are from the same multiplet, errors arising from the 
oscillator strengths and non-LTE effects should be negligible and this 
method has been used previously by, for example,   Dufton et al. 
(\citealp{Duf05}) and Vrancken et al. (\citealp{Vra97}, \citealp{Vra00}).
For two stars (HD\,105071, HD\,125288), the Si\,{\sc iii} lines were too weak to
usefully constrain the microturbulence. For these cases a value of 20 \kms\ was
adopted, which is compatible with the estimates for other supergiants of similar
spectral type.

\subsection{Nitrogen abundances} \label{s_n}

In early B-type stars, there is a relatively rich N {\sc ii} spectra with
singlet transitions at 3995 and 4447 \AA\ and triplet multiplets at 4601-4643,
4780-4819 and 4994-5007 \AA. Here we have estimated nitrogen abundances
solely from the feature at 3995 \AA. This choice was based on the following
considerations:
\begin{enumerate}
\item this feature is amongst the strongest N {\sc ii} lines in the optical spectrum and
is unblended. As such it could be measured in effectively all our targets (see
Table \ref{eq_widths}).
\item the other lines and multiplets were either not observable over the full
range of spectral and/or suffered from blending.
\item the use of only one feature will inevitably increase the random error in
the estimate and may lead to systematic errors. In the case of the former, 
the abundance estimates listed in the Table \ref{Vsini_atm} have a range
of over 1 dex, which is larger than our estimated random error discussed below.
Additionally the nitrogen abundances will depend on the amount of
nucleosynthetically  processed material mixed from the core to the photosphere.
Hence the relative rather than the absolute abundances will be important for
investigating this phenomena. 
\end{enumerate}

The nitrogen abundance estimates form the line at 3995 \AA\ were obtained 
using our grid of {\sc tlusty} models and the methodology discussed, for 
example, in Hunter et al. (\citealp{Hun07}).

To investigate the magnitude of the errors in these abundances, we have also
deduced nitrogen abundance from the other N {\sc ii} lines listed above for
six stars covering the range of spectral types (and hence effective temperatures
present in our sample). These are summarized in Table \ref{N_ab}, where to ease
comparison we have also listed the estimates obtained from the N {\sc ii} line at
3995\AA, which were taken directly from Table \ref{Vsini_atm}. As can be seen there is no systematic differences between the
two set of estimates with the differences being compatible with the standard
deviations found from the abundances estimates from the other features. Hence we
do not believe that our choice of a single feature has introduced significant
systematic errors. The standard deviations in Table \ref{N_ab} range from
$\pm$0.11-0.17 dex and will represent at least in part the random errors in the
abundance determinations. Also there may be systematic
differences in the estimates from different multiplets which will
contribute to the standard deviations of the corresponding estimates.
Hence we adopt a {\it random error} of $\pm$0.15 for the
estimates in Table \ref{Vsini_atm}. We believe that this
may be conservative as the feature at 3995 \AA\ is stronger and does not suffer
from the blending affecting some other feature. As well as
random errors there could additionally systematic errors due to for example
errors in the atmospheric parameters. These have been extensively discussed by
Hunter et al. (\citealp{Hun07}) and vary depending on the strength of the
feature and on the adopted atmospheric parameters. Typically such errors are of
the order of 0.1-0.2 dex (but on occasions can be larger). Here we will adopt a
typical error in our nitrogen abundance estimates of the order of 0.2 dex.

\begin{table}
\caption[]{Nitrogen abundances estimated form the feature at 3995 \AA\ 
(N$_{3995}$) and from other N {\sc ii} lines (N$_o$) for six targets covering
the range of effective temperatures of our sample. For the latter the standard
deviation ($\sigma$) and number ($n$) of the individual estimates are also
listed.}
\begin{center}
\begin{tabular}{lccccc}
\hline\hline
Star   & \teff & N$_{3995}$ & N$_o$ & $\sigma$ & $n$
\\\hline
HD\,115842 & 24800 & 8.44 & 8.28 & 0.17 & 8
\\
HD\,155985 & 23200 & 8.19 & 8.12 & 0.17 & 12
\\
HD\,108002 & 20200 & 8.19 & 8.12 & 0.14 & 12
\\
HD\,142758 & 17300 & 8.46 & 8.50 & 0.13 & 12
\\
HD\,79186  & 15100 & 8.39 & 8.43 & 0.11 & 12
\\
HD\,80558  & 13000 & 8.55 & 8.54 & 0.14 & 12
\\
\hline
\label{N_ab}
\end{tabular}
\end{center}
\end{table}

\subsection{Masses} \label{mass}

Masses for the stars in our sample were estimated by comparing our effective
temperatures and surface gravities estimates with those predicted from 
the evolutonary tracks from the Geneva group models (Meynet \& Maeder, \citealp{Mey03}). 
All the stars in our sample were found to lie in a region between approximately
10\msun\ and 40\msun.

\begin{figure}
\includegraphics[angle=270,width=3.2in]{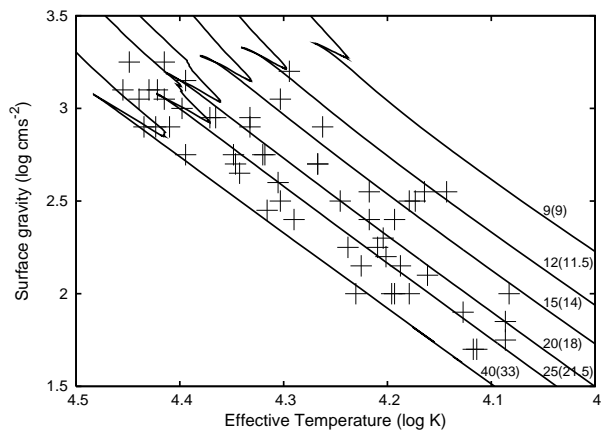}
\caption{Estimated effective temperatures and surface gravities for our sample, 
plotted against the Geneva group evolutionary tracks (Meynet \& Mader, 
\citealp{Mey03}). Each track is identified by an integer which is the initial mass
in solar masses (\msun). The number in brackets refers to the mass of the model at
an effective temperature \teff =  4.1, note that the mass of the models changes 
by less than 0.2 \msun\ between the loop back at the upper left and this point.}
\label{Fig_mass}
\end{figure}

As can be seen from Fig. \ref{Fig_mass}, the evolutionary tracks in our 
pseudo-HR diagram loop back on themselves for values of log \teff\ greater than
4.3 dex and \logg\ greater than 2.8 dex.  This increases the uncertainty in the
determination of mass for stars in this region of the diagram, as a degeneracy
in \logg\ vs \teff\ is introduced. This uncertainty is mitigated by the fact
that the stars in this region all have masses between 20\msun\ and 40\msun\ and
the behaviour of evolutionary models in this mass range is quite similar.

At higher temperatures and lower gravities, the tracks are closer together,
leading to larger errors in the mass estimates, with the relative
error being approximately fifteen per cent. Compounding this problem, stars with high
rotational velocities are likely to exhibit gravity darkening in accordance with
Von Zeipel's Theorem (Von Zeipel \citealp{Von24a}, \citealp{Von24b}), 
and if these stars are pole on, significant errors can be
made in estimating their effective temperature (Gillich, \citealp{Gil08}). These
uncertainties in effective temperature will in turn increase the uncertainty in
mass. Based on these factors, we have adopted a conservative uncertainty in
mass of $\pm $ 20 per cent, which corresponds to an uncertainty of 2\msun\ at the lower
end of our mass range and 8\msun\ at the upper end.

\section{Results and discussion} \label{results}

\subsection{Comparison with previous rotational velocity estimates} \label{comp}

There are several studies of projected rotational velocities that have targets
in common with our sample (see, for example, Abt et al. \citealp{Abt02}; Penny
\citealp{Pen96}; Sim{\'o}n-D{\'{\i}}az \& Herrero \citealp{Sim07}). It is
encouraging that Sim{\'o}n-D{\'{\i}}az \& Herrero  deduced from a similar
methodology to that used here a value of $73\pm4$ \kms\ for HD167264. This is in
reasonable agreement with our estimate of $83\pm3$ \kms\ (our error estimate for
the mean is based on the standard deviation of the sample and assuming that the
errors are normally distributed). Kaufer, Prinja \& Stahl (\citealp{Kau02}) have found
\vsini\ to be 265 $\pm$ 5 \kms for HD 64760, which is again in good agreement with 
our measured value of 255 \kms. It is also encouraging that Kaufer et al. did not use
the Fourier method to obtain this result, but rather rotationally broadened a model 
spectral line to match the Si{\sc iii} line at 4552 \AA. Unfortunately we had no targets 
in common with the sample of Markova \& Puls (\citealp{Mar08}),
 who also used a Fourier technique to measure rotational and macroturbulent 
 velocities. We note, however, that within their limited sample, the trends in both 
 rotational and macroturbulent velocity mirror that of our targets.

\begin{figure}
\includegraphics[angle=270,width=3.5in]{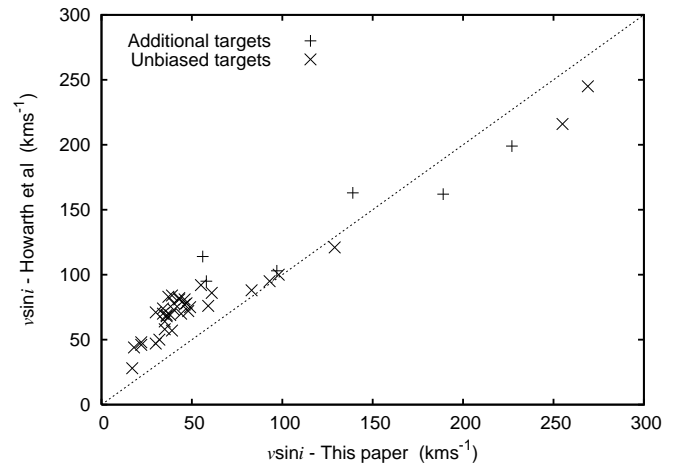}
\caption{Comparison of the projected rotational velocities for stars common 
to both Howarth et al. (\citealp{How97}) and this paper. The effects of 
macrotubulence, which is not explicitly included in the former can be clearly seen.}
\label{vsini}
\end{figure}

The only study that has sufficient overlap to allow a meaningful statistical
comparison is that of Howarth et al. (\citealp{How97}) and in Fig. \ref{vsini} the two sets of
estimates are compared for 42 stars, with spectral types from B0 to B5. The
cross-convolution method of Howarth et al. generally leads to
larger estimates. For the slowly rotating stars (where our estimates are less
than 70 \kms), there is a systematic difference of approximately 30 \kms\ (with
a standard deviation of 10 \kms). We ascribe this difference to the effect of
macroturbulence, which was not explicitly included in the estimates of Howarth
et al. For the stars rotating with intermediate projected velocities between 70
and 140 \kms, we find good agreement between the values  obtained from the cross
correlation and Fourier transform methods; the average difference between the
results of the two techniques is 8 \kms\ with a standard deviation of 8 \kms.
For the four most rapidly rotating stars, HD\,64760, HD\,157246, HD\,99857 and
HD\,93827, the Fourier Tranform estimates are higher by typically 30 \kms\ from
those using the cross-correlation method. We would expect our estimates to be secure as the line
profiles are now dominated by rotational broadening and the zeros in the Fourier
Transform are well observed. It is possible that the difference between our 
values and those of Howarth et al. is due 
to gravity darkening. For rapidly rotating stars, we expect to see a temperature gradient 
across the star, with the poles being hotter than the equator (Von Zeipel \citealp{Von24a}, 
\citealp{Von24b}). This could lead to Howarth et al. finding lower values for \vsini, as their UV
spectra are preferentially sampling the hot, slowly rotating, poles over the cooler, and 
rapidly rotating equator.

\subsection{Comparison with evolutionary models} \label{r_diss}

\begin{figure}
\includegraphics[angle=270,width=3.5in]{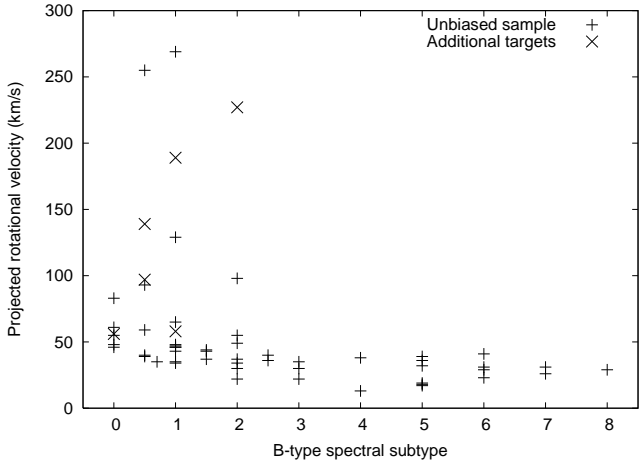}
\caption{Projected rotational velocities as a function of spectral type. The six
additional targets that were selected as they were suspected to have a high
projected rotational velocity have been identified.}
\label{sptype}
\end{figure}

Our estimated projected rotational velocities are plotted against spectral type
in Fig. \ref{sptype}, with the majority of stars lying on a locus from
approximately 60 \kms\ at B0 to 30 \kms\ at B9. This decrease in projected
rotational velocity at later spectral type is in accordance with evolutionary
models, as discussed below.  However, there are a number of early-B type 
supergiants that appear to have anomalously high projected rotational 
velocities for their spectral type and these will be further discussed in Sect.
\ref{r_rapid}. 

We used a Kolmogorov-Smirnov (K-S) test to compare the distribution of rotational
velocities in our Galactic supergiant sample, with that found for supergiants in
the SMC by Dufton et al. (\citealp{Duf06}). While the sample of Dufton et al. is small,
with only 13 stars, the K-S test indicates that at the 0.05 significance level, the distributions
are the same, with a measured D statistic of 0.29. From this, we conclude that there is no evidence for any dependence
of the current rotational velocities on initial metallicity, although this result must be qualified
by again stressing the relatively small sample sizes, and the additional uncertainty
introduced by the unknown value of sin$i$.

%
From a theoretical perspective, the dependance of surface rotational velocity on metallicity is not straightforward. As discussed in Meynet \& Maeder (\citealp{Mey05}), both mass loss and the internal coupling between core and envelope affect the surface equatorial velocity. Lower metallicity reduces mass loss rates (and hence the loss of angular momentum from the star), but it also lowers the efficiency of the transport of angular momentum to the surface. Hence as the metallicity changes, there are two competing mechanisms that to some extent cancel each other out.

Using the models of Meynet \& Maeder (\citealp{Mey03}, \citealp{Mey05}) we compared two 25 \msun\ models with initial velocities of 300 \kms\ and Z=0.02 and Z=0.04 (D25z20S3A and D25z40S3 respectively). The higher metallicity model was found to have a rotational velocity that was $\sim$ 15 \kms\ lower at log \teff\ = 4.35, and $\sim$ 5 \kms\ lower at log \teff\ = 4.1. Such small differences are consistent with the cumulative frequency plot discussed above. The 40 \msun\ models (D40z08S3, D40z20S3A and D40z40S3) show a larger spread of equatorial velocities for a given temperature. For example at a temperature of log \teff\ = 4.2, the low metallicity model (Z=0.008) has v = 37 \kms, the solar metallicity model has v = 43 \kms, and the high metallicity model (Z=0.040) has v = 66 \kms. Scaling these values to account for average projection, the spread in projected rotational velocities is $\sim$ 20 \kms. From this, we conclude that while differences in metallicity may increase the scatter of measured values when compared with evolutionary models, these are unlikely to be large enough to be observable when comparing the distribution and mean of rotational velocites for the SMC B supergiants of Dufton et al. (\citealp{Duf06}) to our Galactic sample. Furthermore, we can discount different metallicities as an explanation for our most rapidly rotating stars, as discussed in Section \ref{r_rapid}.

\begin{figure*}
\centering
\subfigure[9\msun\ to 15\msun]{
\includegraphics[angle=270,width=0.47\textwidth]{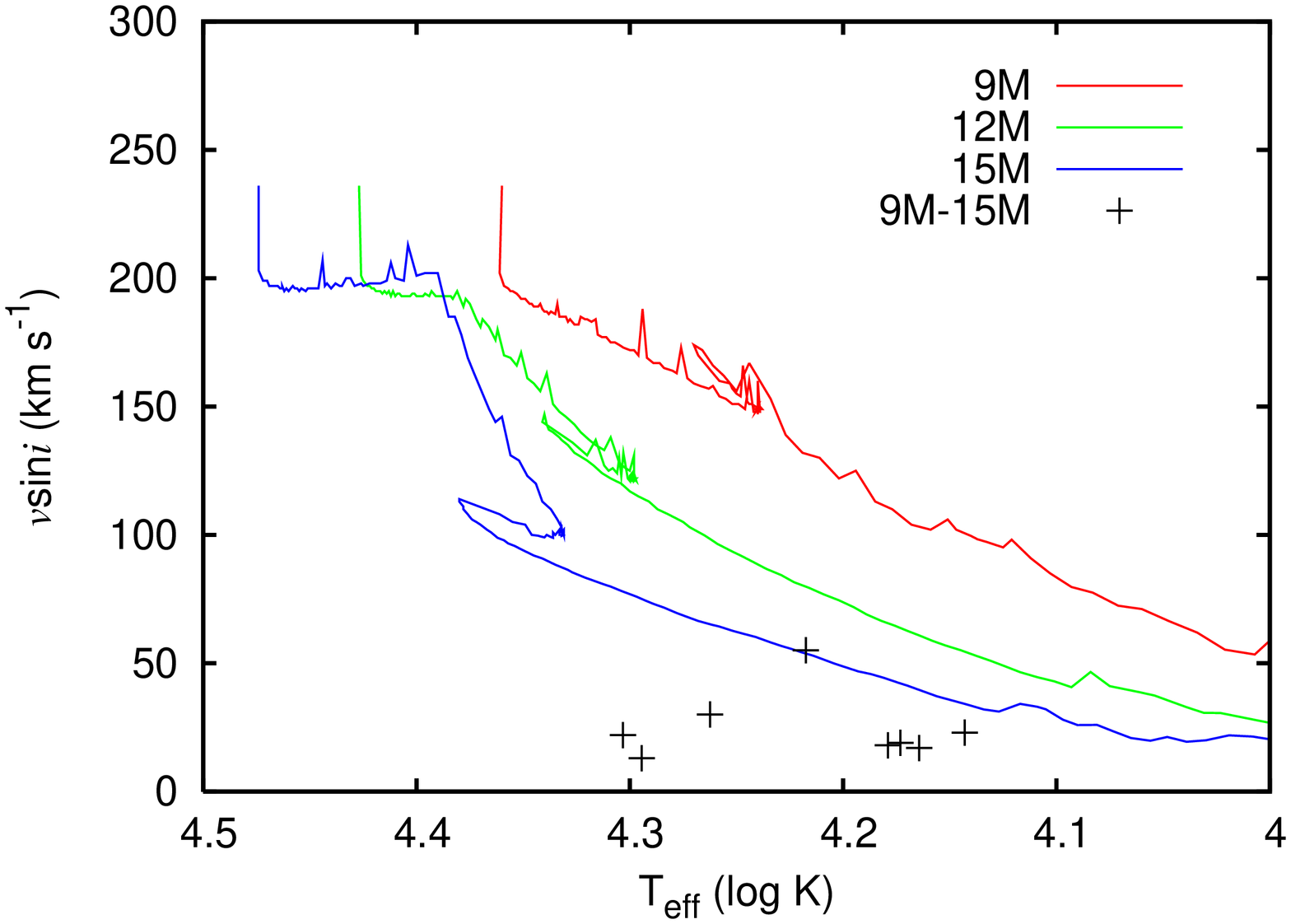}
}
\subfigure[16\msun\ to 20\msun]{
\includegraphics[angle=270,width=0.47\textwidth]{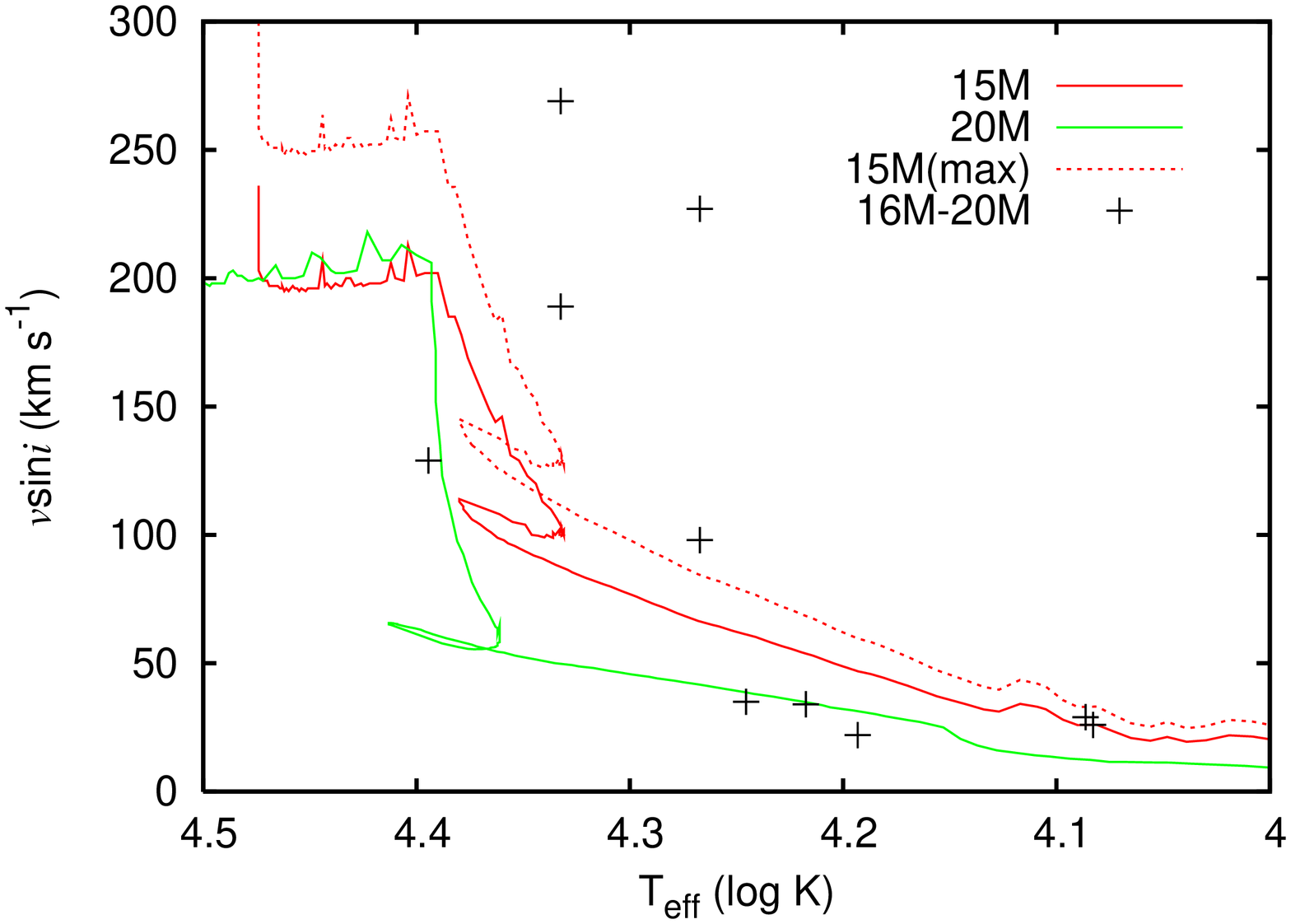}
}
\subfigure[21\msun\ to 25\msun]{
\includegraphics[angle=270,width=0.47\textwidth]{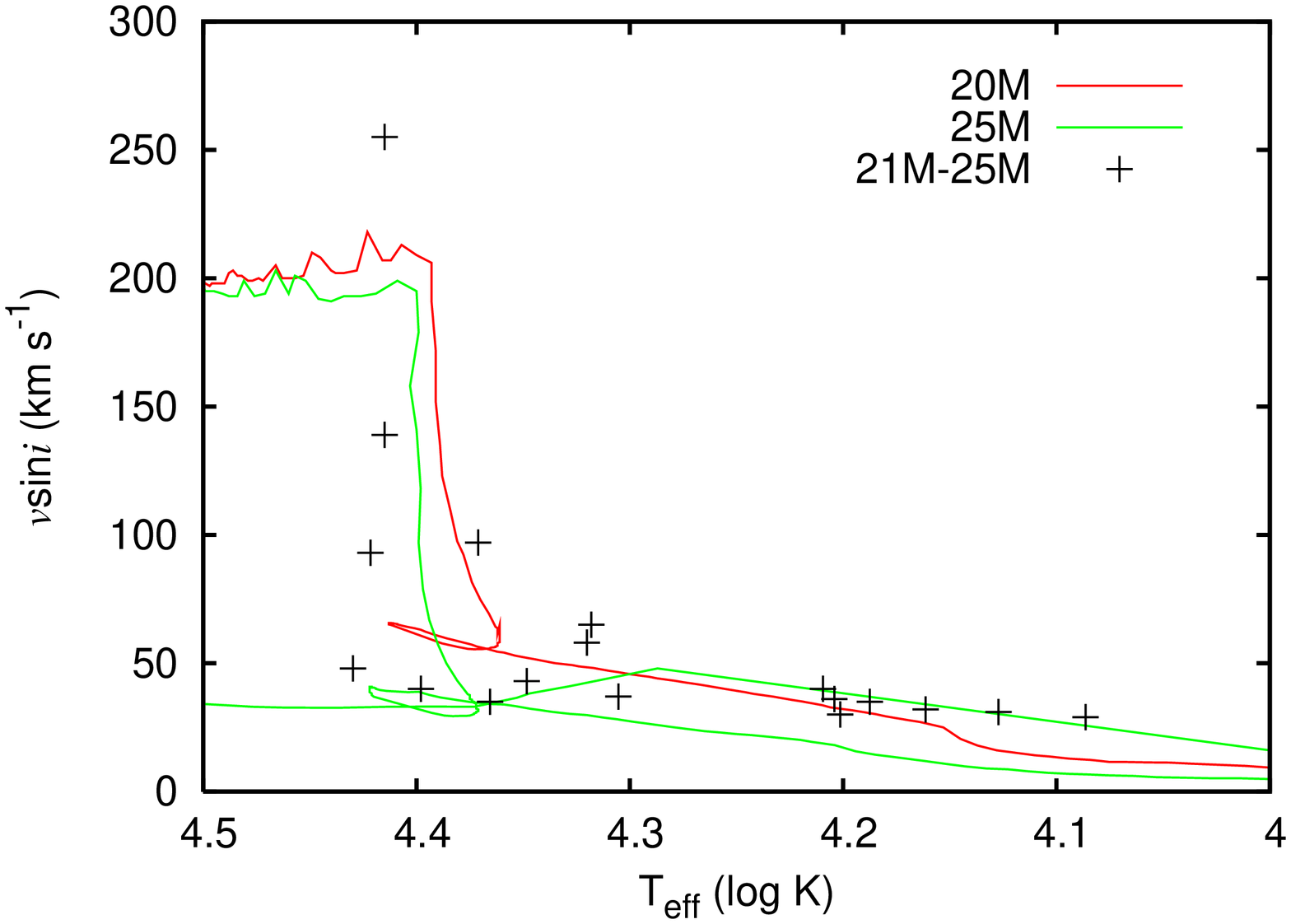}
}
\subfigure[26\msun\ to 40\msun]{
\includegraphics[angle=270,width=0.47\textwidth]{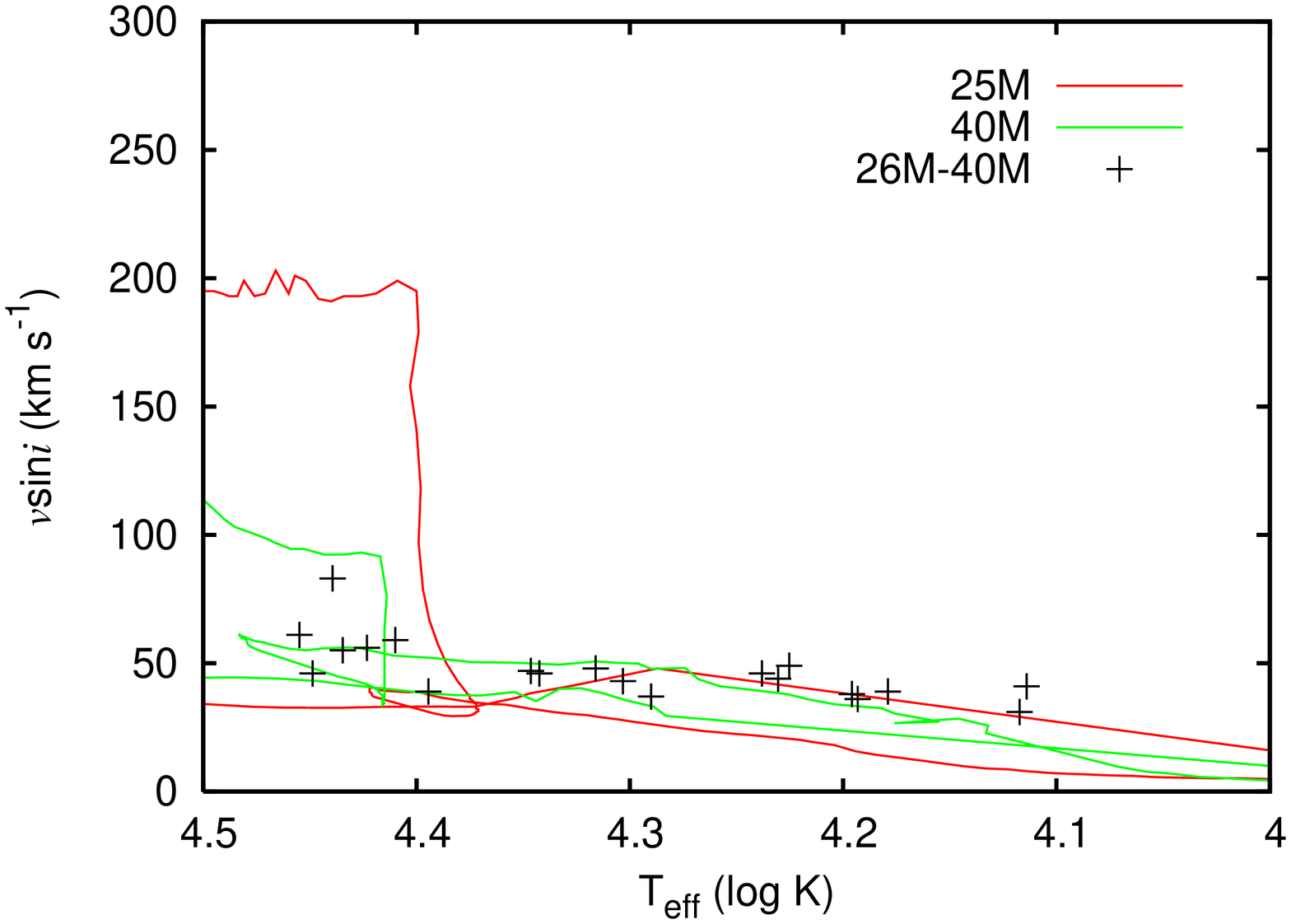}
}
\caption{Projected rotational velocities and effective temperatures for all
stars, plotted against the evolutionary tracks of Meynet and Mader
(\citealp{Mey03}). Note that the rotational velocities from the evolutionary tracks have
been scaled by $\pi/4$ to account for projection, apart from the red
dotted line in the second panel that displays the evolution of a star with an
initial mass of 15 \msun. Stars (which have had their
masses estimated as described in Section \ref{mass}) have been binned
accordingly into four mass ranges 9\msun $<M<$ 15\msun, 16\msun $<M<$ 20\msun,
21\msun $<M<$ 25\msun and 26\msun $<M<$ 40\msun. These mass bins were chosen so
as to facilitate comparison with the available evolutionary models.}
\label{model}
\end{figure*}

The projected rotational velocities  for all our targets  have been plotted
against their effective temperature estimates (from Table \ref{Vsini_atm})  in
Fig. \ref{model}, together with the predictions of the evolutionary  models of
Meynet \& Maeder (\citealp{Mey03}) for Galactic metallicities. To aid the
comparison, the observational dataset has been sub-divided into 
four initial mass ranges. The models used for
the comparison are D09z20S3A, D12z20S3A, D215z20S3A, D20z20S3A, D25z20S3A and
D40z20S3A, which have Z=0.02, an initial equatorial rotational velocity of
300 \kms\ and masses from 9 \msun\ to 40 \msun; the theoretical  rotational 
velocities have been scaled by a factor of $\pi/4$ to convert them to
an average projected rotational velocity assuming a random orientation of axes
(Chandrasekhar \& M\"{u}nch, \citealp{Cha50}). 

In general there is good agreement between observation and theory, both in the
magnitude of the projected rotational velocities and a decrease as one moves to
lower effective temperature or later spectral type. The agreement is
particularly encouraging for the higher mass stars with initial mass estimates
of $> 20$\, \msun. For the two lower mass cohorts, the agreement is less
convincing. For example, the projected rotational velocity estimates for our
lowest mass sample (9 - 15 \msun) appear to be systematically lower than
predicted. This discrepancy would be ameliorated if the initial stellar masses
had been underestimated but as discussed in Sect. \ref{mass}, these estimates
should be reliable to within 2 \msun. Another possibility is that the projected 
rotational velocities have been systematically underestimated. Aerts et al. (\citealp{Aer09})
simulated the effects of non-radial gravity-mode oscillations on the spectral
line profiles of an early-B type supergiant and found that this could lead to an
underestimate of the rotational velocity found by our methodology. Although this
would again improve agreement with theory, it would not explain why the effect
is only found in the lowest mass cohort. Additionally the simulations indicate
that the underestimation only occurs in some cases and hence do not currently
explain the systematic nature of the discrepancy. It should be noted that the
simulations of Aerts et al. are for one set of atmospheric parameters (\teff =
18\,200 K;  \logg = 3.05) and an extension to other spectral types would be
useful. Hence we conclude that supergiants with initial masses in the range 9 -
15 \msun would appear to have lower rotational velocities than are predicted by
the evolutionary models. For supergiants with initial masses of 16 - 20 \msun,
the agreement is good apart from a small number of targets that appear to be
rotating more rapidly and which are discussed further in Sect. \ref{r_rapid}.

We also note that the decrease in rotational velocities seen at spectral type
B1 corresponds to the position of the bi-stability jump (Vink, de Koter \& Lamers, 
\citealp{Vin99}), where wind properties change from a fast wind with
standard mass loss rates, to a slower wind with higher mass loss rates. It 
has been suggested (Vink, in preparation) that the bistability jump may
be accompanied by bi-stability braking, where increased mass loss at spectral
type B1 removes sufficient angular momentum from the system to account for a
dramatic decrease in rotational velocities. Our results for the 20 to 25 \msun\ cohort, 
as illustrated in Fig. \ref{model}, 
provide some tentative support for this rapid decrease in rotational velocity,
with three or four targets targets appearing to lie in this phase.

\subsection{Rapid rotators} \label{r_rapid}

For the purposes of this discussion, ``rapid rotators" refers to four targets, 
HD157246, HD93827, HD99857 \& HD165024 that have larger projected rotational
velocities than predicted by the evolutionary models (see Fig. \ref{model}). Of
these,  HD165024 should be considered as a marginal case; if its mass estimate
was  reduced from 17 to 12 \msun then it would not be classed as rapidly
rotating. However, it was decided to include this in the sample of rapid
rotators that would be examined in more detail for clues as to a different
evolutionary history. 

It should be noted that these rapidly rotating objects appear to be relatively
rare with only four identified in our complete sample. Indeed six targets were 
pre-selected based on them having anomalously large projected rotational
velocities estimates for their spectral type in the analysis of Howarth et al.
Excluding these stars, there are fifty one targets  in our sample, with two
having measured projected rotational velocities  that appear to be significantly
larger than the mean for their spectral type,  viz. HD157246, HD165024. Thus we
find that approximately  4\% of our sample  are anomalously rapid rotators. This
is broadly consistent with the identification of   only one rapid rotator in the
sample of Dufton et al (\citealp{Duf06}) of  13 SMC supergiants. Note that we
may have underestimated the fraction of supergiants that are rapid rotators, as
some of our sample  could be observed pole-on. However as discussed by Hunter 
et al. (\citealp{Hun07}) the probability of observing such objects is low  due
to the nature of the sine function. For example, only thirteen percent of our
targets will have a value of \sini\ less than 0.5, assuming a random orientation
of the rotation axis. Hence we consider that within the  constraints of our
sample size, our estimate of the fraction  of anomalously rapid rotators among
our targets is reasonable and that it is certainly small.

There are at least three channels by which these stars could have reached 
their current evolutionary state: 

\begin{itemize} 
\item they had extremely high
rotational velocities on the main sequence, and have spun down normally as they
evolved.  
\item they have not spun down as much as most other
stars as they became supergiant stars. 
\item they had a moderate
velocity on the main sequence but have experienced an additional source of
angular momentum that has allowed them to maintain their rotational velocity.
\end{itemize}

The first two channels appear implausible as for the first, a very high
main sequence rotational velocity is required, whilst for the second, it is unclear
what mechanism would have inhibited their spin down. Indeed for models in 
this mass range the increase in mass loss due to the bi-stability jump naturally 
leads to a decrease in the rotational velocity.

This leaves the possibility that these stars have an additional source of
angular momentum. A plausible scenario would involve  binarity and mass transfer
(Meibon, Mathieu \& Srassun, \citealp{Mei07}; Langer et al. \citealp{Lan08, Lan03};
Koenigsberger et al. \citealp{Koe03}). The {\sc SIMBAD} online astronomical database
was checked to see if any of the four rapid rotators were listed as
spectroscopic binaries (see Fig.\ref{Vsini_atm}). None of the targets were so classified, and in the
absence of the time-resolved spectra that could indicate binarity from radial
velocity variations, the spectral lines in the rapid rotators were re-examined
but no evidence was found for any asymmetry that might imply the presence of a
binary companion. This is not in itself evidence that these are single stars, 
as the luminous nature of B-supergiants would make the detection of fainter binary
companions difficult.

In addition to these three scenarios, we must also consider the possibility that the models
used are simply not appropriate for these objects. It would be surprising if this was the case
however, as the agreement between theory and observation is excellent for most
of our sample, and there is nothing particularly unusual about the observed properties
of our rapid rotators (besides their large \vsini) that would lead us to suspect the models are
not applicable.

\subsection{Velocity fields} \label{r_macro}

The atmospheres of early-type supergiants appear to contain significant velocity
fields. These are often characterised as microturbulence (estimated in Sect.
\ref{s_xi}) and macroturbulence (estimated in Sect. \ref{macroturbulence}). The
former is incorporated into the line opacity profile and hence directly 
influences the strength of the metal line, whilst the latter is normally
modelled by including additional broadening via a convolution of the 
theoretical line profile.  Assuming that these effects are indeed caused by
velocity fields, they represent the extremes of velocity fields with different
scale lengths in the stellar atmosphere.

The microturbulent velocities deduced for our sample range from approximately 12 to
24 \kms (with two cases where no reliable estimate was possible). There is some
evidence that the microturbulent velocity may be anti-correlated with the effective
temperature but this result should be treated with caution for at least two
reasons. Even with very high quality data, there are significant random errors
associated with the estimation of the microturbulence. For example, the estimate of
the microturbulence for HD\,150898 is 20 \kms. However arbitrarily decreasing the
equivalent width of {Si}{\sc iii} line at 4552 \AA\ by 5 per cent, whilst increasing that
for the line at 4575 \AA\ by 5 per cent would decrease the estimate to 15 \kms.
Additionally the microturbulence estimate can vary depending on the ionic species
and the set of lines considered (see, for example, Vrancken et al. \citealp{Vra97,
Vra00}; Sim\'on-D\'iaz et al. \citealp{Sim06}; Hunter et al. \citealp{Hun07}).
Indeed Vrancken et al. reported differences of up to 9 \kms\ in  the estimates
obtained from different species. Additionally the strength of  the Si {\sc iii}
triplet varies with spectral type leading to different degrees of saturation and
hence of sensitivity to the microturbulence parameter.

Recently Cantiello et al. (\citealp{Can09}) have discussed the obeservational
consequences of sub-surface convection zones in massive stars. These zones arise
from peaks in the opacity (as a function of temperature) that arise from the
ionization of helium and iron. The latter is found to be more prominent at higher
luminosity and lower gravities corresponding in the HR-diagram to the region
occupied by O and B-type supergiants. Using the analyses of Trundle et al.
(\citealp{Tru07}) and Hunter et al. (\citealp{Hun08a}) of B-type stars observed in
the {\sc eso vlt-flames} survey of massive stars (Evans et al. \citealp{Eva05}),
Cantiello et al. find a correlation between the magnitudes of the convection
velocities implied by their models and the estimated microturbulent velocities. In
particular objects with microturbulent velocities of more than 10 \kms\ are in a
region of the HR-diagram where the mean convective velocity within 1.5 pressure
scale heights of the upper boundary of the iron convection zone  ($<v_c>$) is
greater than 2.5 \kms. 

All our targets lie in the region of the HR-digram where the convection due 
to the iron opacity is pronounced. Indeed as well as lying within the contour for
$<v_c>  = 2.5$ \kms, they all lie within the contour for 3.75 \kms\ which was the
largest mean convection velocity regime considered by Cantiello et al. 
For their Galactic sample, Cantiello et al. only considered nine stars with 
five of them having estimates of the microturbulent velocity in excess of 
10 \kms. For our sample, we have been able to estimate the microturbulent
velocity in 54 (out of 57) targets. The large values deduced coupled with their
position within the HR-diagram supports the conclusion of Cantiello et al. that
there is a physical connection between sub-photospheric convection and the small
scale stochastic velocity fields in early-type stars.

As discussed in Sect.\ \ref{macroturbulence}, a macroturbulent velocity was
introduced in order to allow for the excess broadening in the observational
metal line profiles as has been found in other studies (see, for example,
Howarth et al. \citealp{How97}; Ryans et al. \citealp{Rya02}). The velocity
field was assumed to have a Gaussian distribution and in general this appeared
to be consistent with the observations. Estimates were obtained for all but  six
of our targets. In four cases (HD\,64760, HD\,152667, HD\,157246, HD\,93827)
the  line profile is dominated by rotational broadening making the estimation of
the relatively small macroturbulent component unreliable. One target, HD\,83183,
appeared to have a very small degree of macroturbulent broadening, possibly due
to its having a luminosity class II. For the sixth target, HD\,77581, the fits
to the spectra were relatively poor and a reliable estimate was therefore not possible.

The errors for the macroturbulence quoted in Table \ref{Vsini_atm} are the 
standard deviations of the estimates from individual features. As such they do
not include systematic errors and implicitly assume that there is a unique
value  for the macroturbulence. A potentially significant source of systematic
error is the choice of the intrinsic absorption line profiles. These were taken
from spectra calculated from our grid of non-LTE {\sc tlusty} model atmospheres 
as described in Sect. \ref{macroturbulence}. Tests using theoretical spectra
from different grid points indicated that uncertainties in the values adopted
for the effective temperature or surface gravity were unlikely to be
significant. However the choice of the microturbulence (the {\sc tlusty} grid
was calculated for values of 0, 5, 10, 20 and 30 \kms) was more important as it
is this parameter that effectively determines the shape of the theoretical line
profile. Again tests showed that for the early-B type stars, the value adopted
for the microturbulence had little effect of the macroturbulence estimates.
This was because the former had values (10-20 \kms) that were significantly
smaller than the latter (typically 60 \kms). In such cases the intrinsic profile is relatively narrow
compared with the macroturbulent broadening and hence its choice is not
critical. However for late B spectral types, the magnitudes of the
microturbulence and macroturbulence become comparable and in this case the 
choice of the intrinsic line profile becomes more important.

For example, the grid point adopted for HD\,51309 had a microturbulence of 20
\kms\ being the nearest value to the observational estimate of 18 \kms. If
instead we had adopted a microturbulence of 10 \kms, our estimate for the
macroturbulence  would have increased from 20$\pm$3 \kms\ to 31$\pm$5 \kms. This
result is not surprising as we are modelling the total broadening of the lines
using  micro- and macroturbulent velocity fields (together with a rotational
component). A decrease or increase in magnitude of one field will inevitable
anti-correlate with the estimate obtained for the other field. However the
uncertainty in the macroturbulence estimates for the later B spectral types may
be larger than that implied by the statistical error estimates quoted in Table
\ref{Vsini_atm}.

\begin{figure}
\includegraphics[angle=270,width=3.5in]{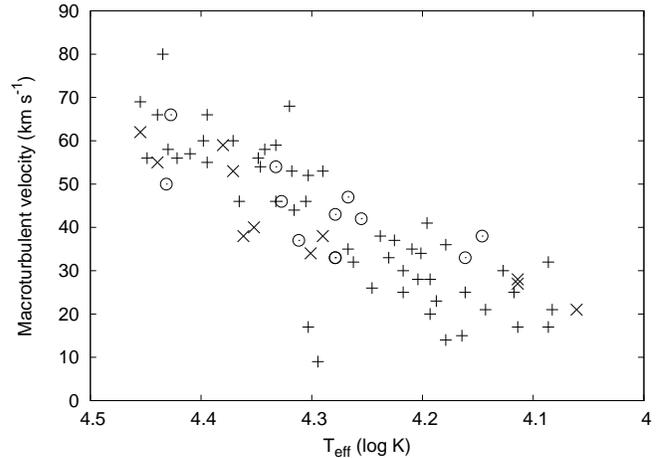}
\caption{Macroturbulent velocity estimates plotted against effective 
temperatures estimates for all stars in our sample for which reliable estimates
were obtained (marked +). Also shown are the estimates of Dufton et al. (\citealp{Duf06})
for Galactic supergiants (marked x) and SMC supergiants (marked o).}
\label{macro_teff}
\end{figure}

The estimates of the macroturbulent velocity show a decrease from approximately
70 \kms at a spectral type of B0 to approximately 20 \kms at a spectral type of
B8. Two targets (HD\,52089 and HD\,159100) appear to have anomalously low
estimates. However these estimates will depend on the adopted  microturbulences
as discussed above. Additionally both these objects have  relatively large
surface gravities for their effective temperature, together with lower estimated
masses than most of our sample; hence their lower estimates may reflect their
different evolutionary status. The decrease of macroturbulence with effective
temperature has been found previously by Ryans et al. (\citealp{Rya02}) for 11
Galactic supergiants and by Dufton et al. (\citealp{Duf06}) for 13 SMC
supergiants. These results have also been plotted on  the Fig. \ref{macro_teff}
and in general the estimates obtained from all three  studies are consistent.

Recently Aerts et al. (\citealp{Aer09}) have provided a physical explanation for
this excess broadening in terms of the non-radial gravity-mode oscillations in 
the stellar atmosphere. They simulate the effect of these oscillations on the
profile of the Si {\sc iii} line at 4552 \AA\ in  a B2 Ia type supergiant  (\teff
= 18\,200 K; \logg = 3.05). Additional broadening is found which is modelled by
rotational broadening coupled with a Gaussian macroturbulent  velocity field.
For the former, the estimates can be lower than those of the models (as
discussed in Sect. \ref{r_diss}), whilst estimates of the macroturbulence
velocity ranging from 0 \kms\ to approximately 50 \kms\ are found. This range is
consistent with our observational estimates in Fig. \ref{macro_teff} and
additionally the range of values found in  the simulations (due to the time
dependent contribution of the different modes)  may also at least partially
explain the scatter found in our estimates, including the anomalously low
estimates found for HD\,52089 and HD\,159100.  

\subsection{Nitrogen abundances} \label{r_nabund} 

Rotation is generally considered to be crucial for developing theoretical 
models of massive star evolution as it can induce mixing of nucleosynthetically
processed material from the stellar core into the photosphere (see, for example,
Heger \& Langer \citealp{Heg00}; Meynet \& Maeder \citealp{Mey00}). Such models
have been used to explain the ratio of blue to red supergiants (Maeder \& Meynet
\citealp{Mae01}) and Wolf-Rayet populations at different metallicities 
(Meynet \& Maeder \citealp{Mey05}; Vink \& de Koter \citealp{Vin05}). A natural
consequence of such mixing is an enhancement of nitrogen (together with smaller
changes in the carbon, oxygen and helium abundances) at the stellar surface.
Hence nitrogen abundances are often used as an indicator of the amount of 
mixing, and in particular rotationally-induced mixing, that has occurred
(see, for example, Brott et al. \citealp{Bro08}).
However recent results from the {\sc VLT-FLAMES} survey (Evans et al. \citealp{Eva05})
have indicated that other mechanisms may affect mixing of material between the
interior and the surface (Hunter et al. \citealp{Hun08b, Hun09}).

\begin{figure}
\includegraphics[angle=270,width=3.5in]{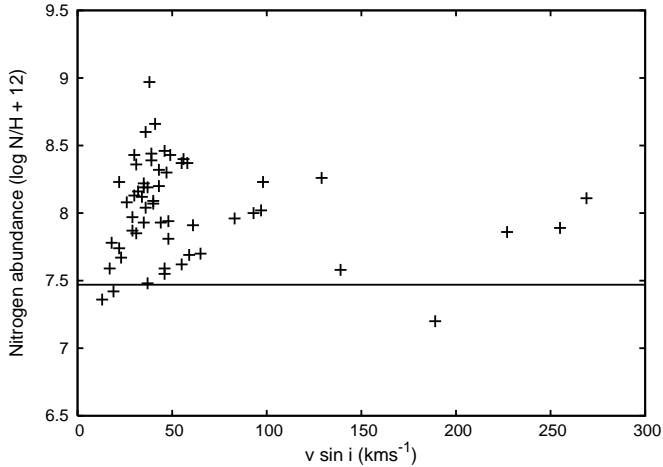}
\caption{Surface nitrogen abundaces versus projected rotational velocity. 
Solid line is base abundance for early B stars while on the main sequence, 
based on VLT-FLAMES data as discussed in text.}
\label{nabund}
\end{figure}

We have estimated the surface nitrogen enhancement for our targets by comparing
our nitrgen abundances with those found for Galactic B-type stars in the
clusters NGC 3293 and NGC 4755 (Trundle et al. \citealp{Tru07}). This study was
choosen as it used the same grid of model atmospheres and
methodology to that adopted here. To ensure  consistency with our own results,
we have only used their abundance estimates from the 3995 \AA\ line, whilst to minimise the
effects of mixing, we have  only included stars of luminosity class V-III. A
mean abundance of  7.47$\pm$0.14 dex was obtained from a sample of 15 stars. 
This is slightly smaller than the mean abundance of 7.59$\pm$0.19 dex   found
when all the nitrogen lines are considered but is consistent with the comparison
of nitrogen abundances given in Table \ref{N_ab}. In Fig. \ref{nabund}, the
surface nitrogen abundance for our targets is plotted  against projected
rotational velocity, together with the base abundance deduced from the results
of Trundle et al.

There would appear to be no correlation of nitrogen abundance with current 
projected rotational velocity. This is not surprising given the ambiguity
arising from the unknown angle of inclination and more importantly the large
change in rotational velocity that occurs when a star leaves the main sequence
(as is apparant, for example, in the theoretical models shown in Fig. 
\ref{model}). We have attempted to allow for the latter effect by estimating the
projected rotational velocity of the progenitor main sequence object. We have
assumed that all the main sequence precursors have a typical surface gravity of 4.2 dex
(Cox \citealp{Cox00});
we note that the actual value adopted is not critical to our methodology. We
have then assumed that the decrease in rotational velocity as a star leaves the
main sequence is due solely to its increase in radius. Hence we are assuming
that the effects of mass loss in both decreasing the stellar moment of inertia 
and in removing angular momentum are negligible. In such circumstances the 
radius 
will simply scale as the square root of the surface gravity and we can estimate the 
ratio of the main sequence to the current radius. Additionally, if we assume 
solid body rotation and that the stellar moment of inertia scales as the 
radius squared, we can
estimate the ratio of the main sequence to the current projected rotational 
velocities and hence deduce estimates for the former. 

To test the validity of our approach, we considered the decrease in equatorial 
rotational velocity as the radius of a star increases, as predicted by the Geneva stellar 
evolutionary models. These tests indicated that our simplification was reasonable, 
with best agreement for the 20\msun\ model. For this model, 
evolving from the main sequence 
(defined here as when the star has a surface gravity of 4.1 dex) to the 
supergiant phase when the star has a surface gravity of 2.9 dex, its 
rotational velocity decreases by a factor of 3.9, whilst its radius increases by 
a factor of 3.8. For more massive models, the agreement is less satisfactory, 
with a 25 \msun\ model increasing in radius by a factor of 
3.8 as it moves from the main sequence to the supergiant stage, but decreasing its 
rotational velocity by a factor of 5.6. At lower masses the agreement is poorer, with 
a 15 \msun\ model increasing its radius by a factor of 4, while decreasing its
rotational velocity by a factor of 2.7. These discrepancies are most likely due to mass loss, 
which we have ignored in our simplification. Furthermore, we have assumed solid body rotation,
whereas in the more sophisticated Geneva models, the coupling between core and envelope and
the changing size of the core will affect the transport of angular momentum within the star, 
and hence rotational velocity.

\begin{figure}
\includegraphics[angle=0,width=3.5in]{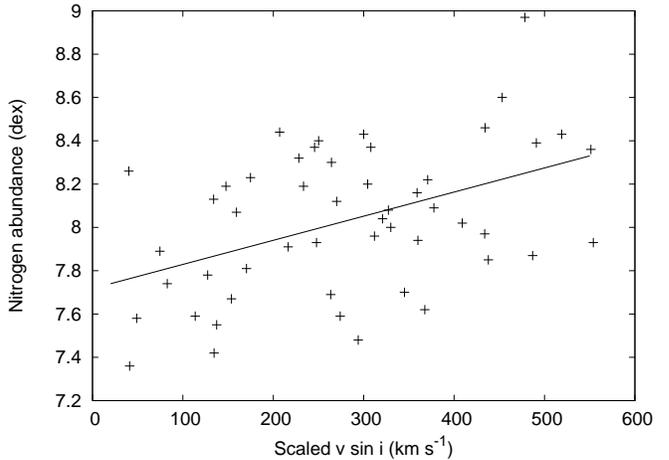}
\caption{Surface nitrogen abundaces versus estimated main sequence 
projected rotational velocity. Also shown is a linear least squares fit.}
\label{nscal}
\end{figure}

In Fig. \ref{nscal}, our nitrogen abundances are plotted against these 
main sequence projected rotational velocities (hencefore designated as
``scaled projected rotational velocities"). We have excluded the four targets
that we believe may not have been undergone normal single star evolution (see
Sect. \ref{r_rapid}) but our conclusions would not be significantly affected if
they were included. We also excluded HD 152236 as its surface gravity 
could not be measured, and HD 64760 as its scaled rotational velocity 
is unrealistically large. We emphasize that the scaled projected rotational
velocities must be considered as subject to significant uncertainties given the 
approximations made. Additionally they do not remove the ambiguity introduced 
by the unknown angle of inclination. However we believe that they are useful as 
the large increase in radius is likely to be an important mechanism as the star 
evolves from the main sequence. 

A correlation is now seen between the scaled projected rotational 
velocity and nitrogen abundance. A linear least squares fit (shown in Fig.
\ref{model} as a solid line) implies an increase in nitrogen abundances of
approximately 0.6 dex (i.e. a factor of four) over our range of scaled projected 
rotational velocities. The slope of the fit is found to be $(1.1 \pm 0.3) \times 10^{-3} $
dex / km s$^{-1}$, which is not consistent with a null result.
Additionally the asymptotic nitrogen abundance at low velocities of approximately
7.7 dex is in reasonable agreement with our baseline nitrogen abundance of 7.47
dex. Indeed as some of the stars with low projected rotational velocities will
have low angles of inclination (and hence relatively large projected velocities)
a small positive offset from the baseline abundance would be expected.
We used a Monte-Carlo technique to try and test the validity of the correlation. The value of
 \vsini\ was randomly varied within one standard deviation of the mean for each star,
and nitrogen abundance and surface gravity varied by 0.2 and 0.1 dex respectively,
and a least squares fit made to the data. For all iterations, a positive correlation 
between scaled \vsini\ and nitrogen abundance was observed, with a nitrogen
enhancement of between 0.3 and 0.6 dex. Based on these results, and within
the limits of our approximations, there would appear to be a statistically significant
correlation between inferred main sequence rotational velocity and surface nitrogen 
abundance at the supergiant stage.

These results are consistent with stellar evolutionary
calculations. For example, Heger \& Langer (\citealp{Heg00}) discussed the
evolution of a star with an initial mass of 20 \msun\ and an equatorial
velocity of 206 \kms. They find an increase of a factor of two or more in the
surface nitrogen abundance at the end of core hydrogen burning.  The Geneva
group stellar evolution models  (Meynet \& Maeder \citealp{Mey00}) also find
similar nitrogen enhancements at the surface of rapidly rotating massive stars.
For example, a 20 \msun\ model rotating with an initial velocity of  200 \kms,
leads to an increase in the nitrogen to carbon ratio at the surface by a factor of
four; when the initial rotational velocity is increased to 300 \kms, this
increases to a factor of approximately seven with most of the change arising from
an enhanced nitrogen abundance. Given the uncertainties in the initial
rotational velocities of our objects, we conclude that there is satisfactory
agreement between our results and models of rotational mixing. However our
results cannot exclude the types of inconsistencies found by Hunter et
al. (\citealp{Hun08b}) in their study of LMC B-type stars.

Although we have compared our results with single star evolutionary models, we
note that in some cases the surface composition may also have been affected by
binary star evolution. For example, while attempting to model the binary
progenitor of SN 1993J, Stancliffe \& Elridge (\citealp{Sta09}) found that a blue
supergiant in a binary system could undergo a surface nitrogen enrichment of up
to a factor of ten.

\subsection{Differential Rotation} \label{r_diffrot}

The relatively simple Fourier transform analysis employed in this study is 
based on the
assumption that stars rotate as a solid body. While observations of our own sun
have long demonstrated this to be false (eg. Schou et al. \citealp{Sch98}), it 
nonetheless remains an appropriate
assumption when confronted with spectra of limited resolution and
signal-to-noise. Reiners \& Schmitt (\citealp{Rei02}) have modeled the
dependence of the first (\(q_1\)) and second (\(q_2\)) zeroes in the 
Fourier transform of
rotationally broadened spectral lines, and have established criteria for solar
and anti-solar like differential rotation. Using this technique, differential
rotation has been identified in A-type main sequence stars by Reiners \& Royer
(\citealp{Rei03}).

Differential rotation has not, however, previously been observed in stars of
comparable spectral type and luminosity class to our sample. We have therefore
attempted to identify the signatures of differential rotation in our sample.
Only the fast rotating stars were considered, as a large projected rotational
velocity allows the reliable identification of more than one zero in the Fourier
Transform. Furthermore, the angle of inclination of the most rapidly rotating
stars can be constrained, as {\em v} $\simeq$ \vsini\ when $i\simeq 90\deg $.
This is important as pole-on rotation may mimic the characteristics of
differential rotation in the Fourier transform of absorption lines.

For the majority of the targets, the second zero was indistinguishable from the
high frequency noise in the transform. However, for five targets (HD152667 
(\(\bar q_{2}/{\bar q_1} = 1.77\)), HD64760 (\(\bar q_{2}/{\bar q_1} = 1.75\)), HD139518 
(\(\bar q_{2}/{\bar q_1} = 1.81\)),  HD100841 (\(\bar q_{2}/{\bar q_1} = 1.74\)) 
and HD157246 (\(\bar q_{2}/{\bar q_1} = 1.87 \pm 0.08\))), the position of the second zero was measurable. Reiners \& Schmitt give  a value of \(q_{2}/{q_1} > 1.83\) as being indicative of anti
solar-like differential rotation (where the poles rotate faster than the
equator), and a value of \(q_{2}/{q_1} < 1.72\) for solar-like differential rotation. 
Hence of these stars, only HD157246 satisfied the criteria for differential rotation, 
and in Table \ref{157246} we summarize the measurements 
for this star. It is important to note however, that the value of \(q_{2}/{q_1}\) 
determined for HD 157246 is within a standard deviation of the threshold for 
differential rotation.

While anti solar-like differential rotation has recently been observed by, for example,
Strassmeir et al. (\citealp{Str03}), from Doppler maps of a K2 giant, it is
unclear what the physical mechanism behind this is. It has been proposed that
meridional circulation, driven by a strong temperature gradient, could lead to
anti solar-like differential rotation (Kichatinov \& R\"{u}diger,
\citealp{Kic05}).

\begin{center}
\begin{table}
\caption[]{Details of lines used for analysis of differential rotation in HD157246}
\begin{tabular}{lccccc}\hline\hline
Species   & $\lambda$ (\AA) & \vsini & \(q_1\) & \(q_2\) & $\frac{q_2}{q_1}$\ 
\\\hline
He I   & 3819.61           & 264 & 0.194 & 0.359 & 1.85 
\\
N II   & 3995.00           & 279 & 0.177 & 0.335 & 1.89
\\
He I   & 4026.19           & 269 & 0.181 & 0.346 & 1.91 
\\	
C II   & 4267.00 / .26 & 255 & 0.181 & 0.335 & 1.85 
\\
Mg II  & 4481.13 / .33 & 255 & 0.171 & 0.296 & 1.73 
\\
Si III & 4552.62           & 277 & 0.156 & 0.296 & 1.90 
\\
O II   & 4661.63           & 279$^*$ & 0.150 & 0.301 & 2.01 
\\
He I   & 4921.93           & 271 & 0.148 & 0.274 & 1.85 
\\
\hline
\multicolumn{6}{l}{\(^*\) Poor quality line, hence \(q_1\) and \(q_2\) are uncertain.}
\label{157246}
\end{tabular}
\end{table}
\end{center}

\section{Conclusion} \label{conc}

From a sample of more than fifty early type supergiant stars, we find  a range
of rotational velocities that in general are in good agreement with evolutionary
models. For four stars, we find a significant difference between their measured
values and the expected rotational velocities for their current evolutionary
status. These discrepancies cannot be explained in terms of projection effects
and are  unlikely to be caused by observational uncertainties, and we conclude
that the most likely explanation for their behavior is binarity. In addition to
the projected rotational velocities, we have measured the macroturbulence and
microturbulence in most of our targets. For the former, we find good agreement
with the predictions of Cantiello et al. (\citealp{Can09}) for convection 
driven by the opacity of iron group elements. The latter may be a manifestation
of the large number of  non-radial gravity-mode stellar oscillations (Aerts et
al. \citealp{Aer09}) present in the stellar photosphere and our results are
consistent with their prediction for a B2 Ia supergiant.

We find a range of nitrogen abundances stetching from the Galactic B-type
baseline abundance to an enhancement of more than 1 dex. No correlation of
abundance is found with the current projected rotational velocity. However a
correlation is found with the estimated main sequence projected rotational
velocity and this is consistent with current evolutionary models for single
rotating stars. 

We have attempted to determine if there was any evidence of
differential rotation in our sample, and find marginal evidence that one of our most 
rapidly rotating targets is undergoing anti-solar like differential rotation.

\section{Acknowledgements} \label{acc}

The authors thank Connie Aerts for her comments and advice on an early draft of
this paper, and to Chris Evans for obtaining some preliminary observational data. 
We also thank Ines Brott and Jorick Vink for useful discussions. We thank the referee, Ian Howarth,
for his insightful comments, and in particular the suggestion that gravity darkening may explain
the discrepancies between the values of \vsini\ found for rapidly rotating stars in this work,
and with cross-correlation methods. This research has made use
of NASA's Astrophysics Data System and the SIMBAD database, operated at CDS, 
Strasbourg, France. This research was supported by a rolling grant 
awarded by the UK Science and Technology Facilities Council. MF is funded by the 
Northern Ireland Department of Employment and Learning.

\bsp

\label{lastpage}

\end{document}